\documentclass[aps,prd,reprint,amsmath,amssymb,nofootinbib,longbibliography,superscriptaddress]{revtex4-2}
\usepackage[T1]{fontenc}
\usepackage{microtype}
\usepackage{hyperref}
\allowdisplaybreaks[1]
\hypersetup{colorlinks=true,linkcolor=blue,citecolor=blue,urlcolor=blue,pdftitle={Relativistic Covariance from Hamiltonian Hydrodynamics in the Chiral Limit}}
\begin{document}
\title{Relativistic Covariance from Hamiltonian Hydrodynamics in the Chiral Limit}

\author{Zi-Hao Liu}
\email{zihaomology@gmail.com}
\affiliation{
School of Physical Sciences, University of Chinese Academy of Sciences,
Beijing 101408, China
}

\author{Mei Huang}
\affiliation{
School of Nuclear Science and Technology, University of Chinese Academy of Sciences,
Beijing 101408, China
}

\date{September 18, 2026}

\begin{abstract}
We derive the covariant equations of nondissipative chiral hydrodynamics from their equal-time Hamiltonian formulation and identify the Lie--Poisson structure implicit in the chiral brackets. This structure combines fluid motion with local chiral transformations. For a parity-invariant Hamiltonian quadratic in chiral charge densities and pion gradients, without external backgrounds, we derive and solve the constraints imposed by Lorentz boosts. Their coupled nonlinear sector is equivalent to the linear boost transformation laws of a symmetric tensor, which determine the velocity dependence from rest-frame data. For finite, positive rest-frame coefficients, finite ordinary-fluid energy density and thermodynamic derivatives, and a momentum--velocity relation that can be solved in either direction, the chiral Hamiltonian coefficients remain finite at every fluid speed below light speed if and only if the rest-frame pion propagation speed does not exceed the speed of light. This formulation makes symmetry constraints explicit and provides a starting point for future thermodynamically consistent dissipative extensions.

\end{abstract}

\maketitle

\section{Introduction}

In the chiral limit, pion fields in the chirally broken phase of QCD evolve together with the conserved fluid densities~\cite{Son2000,Son2001}. The Hamiltonian approach encodes their coupled dynamics in the energy and charge algebra, providing a direct way to impose symmetry. Ref.~\cite{Son2000} stated the boost condition and covariant equations, while Ref.~\cite{Pujol2003} quoted the velocity-dependent coefficients. The central question is how the boost condition constrains the equal-time Hamiltonian and whether it uniquely fixes the velocity dependence from rest-frame data. We establish this connection for the parity-invariant theory without external backgrounds, at quadratic order in chiral charge densities and pion gradients. This distinguishes the velocity dependence required by relativity from the rest-frame functions supplied by thermodynamics or microscopic theory.

We identify the Lie--Poisson structure underlying the chiral brackets to specify which transformations are encoded in the equal-time algebra. Spatial transformations act on the local chiral current algebra through a semidirect product. Including their action on the fluid densities and pion field reproduces the full Poisson bracket and explains its Jacobi identity~\cite{Marsden1984,Holm1998,Morrison1998}. This structure separates the local transformations generated by the brackets from the global chiral symmetry required of the Hamiltonian.

Energy conservation determines the energy flux, while boost invariance requires that flux to equal the independent momentum density~\cite{Dirac1949,Dirac1962,Schwinger1963}. We show that the resulting three coupled nonlinear constraints are equivalent to the linear boost transformation laws of a symmetric tensor. Together with the independent vector-charge constraint, this equivalence uniquely fixes the velocity dependence from three rest-frame functions for all fluid velocities reached from rest before any coefficient diverges. The tensor structure follows from the Hamiltonian constraints themselves and yields the covariant pion equation, chiral currents, stress tensor, and thermodynamics.

The same tensor relates the Hamiltonian coefficients to pion propagation. We assume finite ordinary-fluid energy density and thermodynamic derivatives, with a momentum--velocity relation that can be solved in either direction. For finite, positive rest-frame coefficients, the chiral coefficients remain finite at every fluid speed below light speed if and only if the rest-frame pion speed does not exceed light speed. If this bound is violated, the divergence occurs where a pion wavefront lies along the equal-time hypersurface.

The separation of Poisson structure and Hamiltonian also provides a basis for dissipative extensions through GENERIC (General Equation for the Nonequilibrium Reversible--Irreversible Coupling), which combines Hamiltonian evolution with entropy production~\cite{Grmela1997,OttingerGrmela1997}. The present construction supplies the reversible dynamics. A complete dissipative theory additionally requires an operator compatible with energy conservation, nonnegative entropy production, and relativistic symmetry~\cite{Ottinger1998}; its construction remains future work.

\section{Equal-time Hamiltonian formulation}
\subsection{Equal-time Poisson algebra}
We begin with the variables on a fixed equal-time hypersurface. The entropy and baryon-number densities are $s(\mathbf x)$ and $n(\mathbf x)$, and the momentum density $T^{0i}(\mathbf x)$ is an independent Hamiltonian variable. The left- and right-handed flavor-charge densities are $\rho_L^a(\mathbf x)$ and $\rho_R^a(\mathbf x)$. The phase of the chiral order parameter is a field $\Sigma(\mathbf x)\in SU(N_f)$, with $N_f\geq2$, transforming as $\Sigma\to L\Sigma R^\dagger$. The independent variables are therefore
\begin{equation}
s,\quad n,\quad T^{0i},\quad \rho_L^a,\quad \rho_R^a,\quad \Sigma.
\end{equation}
We use natural units, $c=\hbar=1$. Greek indices label spacetime components, $i,j,k,\ldots$ label spatial components, and $a,b,c,\ldots$ label adjoint flavor components. Spatial contractions use $\delta_{ij}$; the covariant metric, introduced later, is $g^{\mu\nu}=\operatorname{diag}(1,-1,-1,-1)$. Traces are taken in the fundamental flavor representation, and $X=L,R$ labels chirality. We write $\sum_{X,a}$ for the sum over both chiralities and all adjoint components.

The Hermitian generators $\lambda^a/2$ define the charge matrices $\rho_{L,R}=\rho_{L,R}^a\lambda^a/2$ and satisfy
\begin{equation}
\operatorname{tr}\!\left(\tfrac{\lambda^a}{2}\tfrac{\lambda^b}{2}\right)
 =\tfrac12\delta^{ab},\quad
[\lambda^a/2,\lambda^b/2]=if^{abc}\lambda^c/2,
\end{equation}
where $f^{abc}$ are the structure constants. We take $\partial_0F=\{H,F\}$. The internal brackets of Ref.~\cite{Son2000} are
\begin{align}
\{\rho_{L,R}^a(\mathbf x),\rho_{L,R}^b(\mathbf y)\}
 &=-f^{abc}\rho_{L,R}^c(\mathbf x)\delta(\mathbf x-\mathbf y),\notag\\
\{\rho_L^a(\mathbf x),\Sigma(\mathbf y)\}
 &=-i\frac{\lambda^a}{2}\Sigma(\mathbf x)\delta(\mathbf x-\mathbf y),\notag\\
\{\rho_R^a(\mathbf x),\Sigma(\mathbf y)\}
 &=i\Sigma(\mathbf x)\frac{\lambda^a}{2}\delta(\mathbf x-\mathbf y).\label{p:chiralPB}
\end{align}
The left and right charge densities Poisson-commute, as do all components of the order-parameter fields. The brackets of $\Sigma^\dagger$ follow by complex conjugation. The spatial brackets are
\begin{align}
\{T^{0i}(\mathbf x),A(\mathbf y)\}
 &=A(\mathbf x)\partial_{x^i}\delta(\mathbf x-\mathbf y),\notag\\
\{T^{0i}(\mathbf x),\Sigma(\mathbf y)\}
 &=-\partial_i\Sigma(\mathbf x)\delta(\mathbf x-\mathbf y),\notag\\
\{T^{0i}(\mathbf x),T^{0j}(\mathbf y)\}
 &=\bigl[T^{0j}(\mathbf x)\partial_{x^i}-T^{0i}(\mathbf y)\partial_{y^j}\bigr]\delta(\mathbf x-\mathbf y),\label{p:spacePB}
\end{align}
Here $A$ denotes any one of the densities $s,n,\rho_L^a,\rho_R^a$. All boundary terms below vanish by periodicity or sufficient decay.

\subsection{Transformations generated by the brackets}
Smearing the momentum and charge densities makes their generated transformations explicit. Define
\begin{align}
P[\xi]&=\int d^3x\,\xi^iT^{0i},\notag\\
Q_L[\theta_L]&=\int d^3x\,\theta_L^a\rho_L^a,\quad
Q_R[\theta_R]=\int d^3x\,\theta_R^a\rho_R^a,\label{p:smeared}
\end{align}
where $\xi^i(\mathbf x)$ is a spatial vector field and $\theta_{L,R}^a(\mathbf x)$ are test functions. The latter are local current-algebra parameters, not hydrodynamic variables. All spatial derivatives of these parameters are taken to exist and be continuous. \eqref{p:chiralPB} and \eqref{p:spacePB} give
\begin{equation}\label{p:smearedaction}
\begin{alignedat}{2}
\{P[\xi],\Sigma\}&=-\xi^i\partial_i\Sigma, &\quad \{P[\xi],A\}&=-\partial_i(\xi^iA),\\
\{Q_L[\theta_L],\Sigma\}&=-i\theta_L\Sigma, &\quad \{Q_R[\theta_R],\Sigma\}&=+i\Sigma\theta_R,
\end{alignedat}
\end{equation}
where $\theta_{L,R}=\theta_{L,R}^a\lambda^a/2$. The first two relations distinguish the spatial scalar $\Sigma$ from the scalar densities $s,n,\rho_L^a,\rho_R^a$. The last two give the infinitesimal left and right chiral rotations.

Because the chiral parameters depend on position, a spatial transformation also acts on the smeared charges. Integration by parts gives
\begin{align}
\{P[\xi],Q_L[\theta_L]\}
 &=-\int d^3x\,\theta_L^a\partial_i(\xi^i\rho_L^a) =Q_L[\xi^i\partial_i\theta_L],
\notag\\\{P[\xi],Q_R[\theta_R]\}
 &=Q_R[\xi^i\partial_i\theta_R].\label{p:mixedPQ}
\end{align}
The remaining brackets close as
\begin{align}
\{Q_L[\theta],Q_L[\phi]\}&=Q_L[i[\theta,\phi]],\notag\\
\{Q_R[\theta],Q_R[\phi]\}&=Q_R[i[\theta,\phi]],\notag\\
\{Q_L[\theta],Q_R[\phi]\}&=0,\notag\\
\{P[\xi_1],P[\xi_2]\}&=P[[\xi_1,\xi_2]],\label{p:smearedclosure}
\end{align}
where $[\xi_1,\xi_2]^i=\xi_1^j\partial_j\xi_2^i-\xi_2^j\partial_j\xi_1^i$. With Hermitian flavor parameters, $i[\theta,\phi]$ is Hermitian and has components $-f^{abc}\theta^a\phi^b$, in agreement with \eqref{p:chiralPB}. The mixed bracket in \eqref{p:mixedPQ} shows that spatial vector fields act on local flavor parameters by differentiation. The momentum and charge generators therefore realize the semidirect-product Lie algebra
\begin{equation}
\mathfrak X(\mathbb R^3)\ltimes
C^\infty\!\left(\mathbb R^3,
\mathfrak{su}(N_f)_L\oplus\mathfrak{su}(N_f)_R\right).
\label{p:semidirectalg}
\end{equation}
The positive sign in \eqref{p:mixedPQ} follows by integrating the density variation $-\partial_i(\xi^i\rho)$ against the fixed test function $\theta$.

The corresponding finite action fixes the composition convention. A diffeomorphism $\varphi$ acts on the scalar $\Sigma$ as $\Sigma(\mathbf x)\mapsto\Sigma(\varphi^{-1}(\mathbf x))$. If $g_L(\mathbf x)$ and $g_R(\mathbf x)$ are space-dependent chiral transformations, then
\begin{equation}
[(\varphi,g_L,g_R)\cdot\Sigma](\mathbf x)
 =g_L(\mathbf x)\Sigma(\varphi^{-1}(\mathbf x))g_R^\dagger(\mathbf x).
\end{equation}
Composing two such transformations gives
\begin{align}
&(\varphi_1,g_{1L},g_{1R})(\varphi_2,g_{2L},g_{2R})\notag\\
&\quad=\Bigl(\varphi_1\circ\varphi_2,
 g_{1L}(g_{2L}\circ\varphi_1^{-1}),
 g_{1R}(g_{2R}\circ\varphi_1^{-1})\Bigr).\label{p:semidirectgroup}
\end{align}
This is the semidirect product
\begin{equation}
\operatorname{Diff}(\mathbb R^3)\ltimes
C^\infty\!\left(\mathbb R^3,SU(N_f)_L\times SU(N_f)_R\right)
\end{equation}
for the stated convention. Semidirect products and their Lie--Poisson reductions are standard in Hamiltonian fluid mechanics~\cite{Marsden1984,Holm1998,Morrison1998,GayBalmaz2009}. Their use for non-Abelian fluids goes back to chromohydrodynamics~\cite{Gibbons1982,HolmKupershmidt1984b}; related constructions appear in Refs.~\cite{Bistrovic2003,Jackiw2004,GayBalmaz2011}. Here they identify the composition law already encoded in the chiral brackets.

These local transformations describe the current algebra. The Hamiltonian below is required to have global $SU(N_f)_L\times SU(N_f)_R$ symmetry; arbitrary local chiral transformations and arbitrary $\xi^i$ need not leave it invariant.

The full bracket inherits the Jacobi identity from a Lie--Poisson construction. Before imposing unitarity, spatial pullback and left and right chiral multiplication act linearly on the real vector space of complex-matrix-valued fields. Together with the density representations for $s$ and $n$, these actions make $(s,n,\Sigma)$ advected variables in the construction of Ref.~\cite{Holm1998}. Extending \eqref{p:semidirectalg} by the Abelian representation space dual to these fields, with the real trace pairing for matrices, gives a Lie--Poisson bracket reproducing \eqref{p:chiralPB} and \eqref{p:spacePB}. Pullback and chiral multiplication preserve $\Sigma\in SU(N_f)$. Since the advected variables Poisson-commute, every Hamiltonian vector field is tangent to this subset. It is therefore a Poisson submanifold, and the restricted bracket satisfies the Jacobi identity.

\subsection{Symmetry constraints on the Hamiltonian}
We retain the full fluid-velocity dependence while making two truncations. The hydrodynamic derivative expansion is restricted to the nondissipative ideal-fluid level~\cite{LandauLifshitz,Kovtun2012}. Separately, the chiral Hamiltonian is kept to quadratic order in $\rho_L,\rho_R$ and the first pion gradients $\partial_i\Sigma,\partial_i\Sigma^\dagger$. It contains no gradients of $s,n,T^{0i}$ or $\rho_{L,R}$. Write its density as $T^{00}$ and its ordinary-fluid part as $\epsilon(s,n,T^{0i})$. Then $T^{00}-\epsilon=O[(\rho_{L,R},\partial\Sigma)^2]$, and products of two such corrections are quartic and are discarded. Neither the momentum density nor the fluid velocity is assumed small.

Assume global $SU(N_f)_L\times SU(N_f)_R$, spatial rotations, parity, Hermitian charge matrices, and real coefficients. There are no external flavor or spatial tensors and no parity-odd background parameters. The chiral transformations are
\begin{equation}
\rho_L\to L\rho_LL^\dagger,\quad \rho_R\to R\rho_RR^\dagger,
\quad\Sigma\to L\Sigma R^\dagger.
\end{equation}
The matrices $\rho_L$, $\Sigma\rho_R\Sigma^\dagger$, and $\Sigma\partial_i\Sigma^\dagger$ transform by the same left conjugation. They are traceless; the trace is the invariant bilinear. Because the chiral group acts transitively on $\Sigma$, a chiral-invariant potential depending only on $\Sigma$ must be constant. The density bilinears reduce to the two squares $\operatorname{tr}\bigl(\rho_L-\Sigma\rho_R\Sigma^\dagger\bigr)^2$, $\operatorname{tr}\bigl(\rho_L+\Sigma\rho_R\Sigma^\dagger\bigr)^2$, and
\begin{equation}
\operatorname{tr}[\bigl(\rho_L-\Sigma\rho_R\Sigma^\dagger\bigr)\bigl(\rho_L+\Sigma\rho_R\Sigma^\dagger\bigr)]=\operatorname{tr}\rho_L^2-\operatorname{tr}\rho_R^2.
\end{equation}
Parity exchanges $\rho_L,\rho_R$, sends $\Sigma(t,\mathbf x)$ to $\Sigma^\dagger(t,-\mathbf x)$, and reverses the momentum density. The last bilinear is odd. No pseudoscalar coefficient can be formed from $s,n$ and the single polar vector $T^{0i}$, so it is excluded.

The gradient trace is real and symmetric:
\begin{align}
\operatorname{tr}(\partial_i\Sigma\partial_j\Sigma^\dagger)
 &=-\operatorname{tr}[(\partial_i\Sigma\Sigma^\dagger)(\partial_j\Sigma\Sigma^\dagger)]\notag\\
 &=\operatorname{tr}(\partial_j\Sigma\partial_i\Sigma^\dagger)
 =[\operatorname{tr}(\partial_i\Sigma\partial_j\Sigma^\dagger)]^*.\label{p:gradreal}
\end{align}
Its coefficient uses $\delta_{ij}$ and $T^{0i}T^{0j}$. An antisymmetric coefficient, including $\epsilon_{ijk}T^{0k}$, contracts to zero. Single-trace linear terms vanish by tracelessness, so double traces give no further quadratic invariant.

Before imposing parity there are two density--gradient traces,
\begin{align}
&\operatorname{tr}(\rho_L\Sigma\partial_i\Sigma^\dagger
           \mathbin{\pm}\Sigma^\dagger\partial_i\Sigma\rho_R)\notag\\
&\quad=\operatorname{tr}[(\rho_L\mathbin{\mp}\Sigma\rho_R\Sigma^\dagger)
                           \Sigma\partial_i\Sigma^\dagger].
\end{align}
Under parity, the trace with the plus sign on the left is a polar vector, while the one with the minus sign is an axial vector. Only the former can contract with a coefficient parallel to $T^{0i}$ to form a parity-even scalar. With no available axial-vector coefficient, the Hamiltonian is
\begin{widetext}
\begin{align}
H&=\int d^3x\,T^{00},\notag\\
T^{00}&=\epsilon+f_{ij}\operatorname{tr}(\partial_i\Sigma\partial_j\Sigma^\dagger)
 +\frac a2\operatorname{tr}\bigl(\rho_L-\Sigma\rho_R\Sigma^\dagger\bigr)^2+\frac{a_v}{2}\operatorname{tr}\bigl(\rho_L+\Sigma\rho_R\Sigma^\dagger\bigr)^2
 -ic_i\operatorname{tr}(\rho_L\Sigma\partial_i\Sigma^\dagger+\Sigma^\dagger\partial_i\Sigma\rho_R),\label{p:H}
\end{align}
\end{widetext}
Here $\epsilon(s,n,T^{0i})$ is the ordinary-fluid energy density obtained by setting the chiral densities and pion gradients to zero. The functions $a,a_v,c_i,f_{ij}$ are state-dependent Hamiltonian coefficients, all depending on $(s,n,T^{0i})$, and $f_{ij}=f_{ji}$. Under the stated symmetry and truncation assumptions, these exhaust the terms in (13) of Ref.~\cite{Son2000}. Terms with additional fluid gradients are not included in this classification.

The mixed term needs no added Hermitian conjugate. Separately,
\begin{align}
[\operatorname{tr}(\rho_L\Sigma\partial_i\Sigma^\dagger)]^*
 &=\operatorname{tr}(\rho_L\partial_i\Sigma\Sigma^\dagger)
 =-\operatorname{tr}(\rho_L\Sigma\partial_i\Sigma^\dagger),\notag\\
[\operatorname{tr}(\Sigma^\dagger\partial_i\Sigma\rho_R)]^*
 &=\operatorname{tr}(\rho_R\partial_i\Sigma^\dagger\Sigma)
 =-\operatorname{tr}(\Sigma^\dagger\partial_i\Sigma\rho_R).\label{p:crossreal}
\end{align}
Cyclicity and differentiated unitarity give both identities. Each trace is purely imaginary, so multiplication by real $-ic_i$ makes each contribution real. Adding the Hermitian conjugate would double the mixed term.

The real polar vector
\begin{equation*}
-i\operatorname{tr}(\rho_L\Sigma\partial_i\Sigma^\dagger
      +\Sigma^\dagger\partial_i\Sigma\rho_R)
\end{equation*}
is bilinear in the charge densities and pion gradients. Contracting two copies of this expression gives a quartic scalar, not an additional quadratic invariant. Contracting it with an external polar vector would introduce a background excluded here. Contracting it instead with $\partial_i n$ is allowed by the symmetries but introduces a fluid gradient, so it is excluded by the ideal-fluid gradient restriction rather than by its order in the chiral variables. Terms with further pion derivatives likewise lie outside the stated truncation.

\section{Momentum and energy currents}
\subsection{Momentum current}
To derive the momentum current, we first specify the matrix-derivative convention. Treat $\Sigma$ and $\Sigma^\dagger$ as separate arguments of \eqref{p:H}, with $(\partial T^{00}/\partial\Sigma)_{ab}=\partial T^{00}/\partial\Sigma_{ba}$. Their variations then enter as $\operatorname{tr}[(\partial T^{00}/\partial\Sigma)\delta\Sigma]$, and similarly for $\Sigma^\dagger$ and the gradient arguments. Physical variations satisfy $\delta\Sigma^\dagger=-\Sigma^\dagger\delta\Sigma\Sigma^\dagger$. Although individual derivatives can depend on the extension away from the unitary constraint, their contractions with physical variations do not; those contractions determine the dynamics.

The Hamiltonian conjugates are
\begin{equation}
T=\frac{\partial T^{00}}{\partial s},\qquad
\mu=\frac{\partial T^{00}}{\partial n},\qquad
v_i=\frac{\partial T^{00}}{\partial T^{0i}}.
\label{p:conjugates}
\end{equation}
Here $T$ and $\mu$ are conjugate to the entropy and baryon densities, and $v_i$ is conjugate to momentum density. Its interpretation as fluid velocity will follow from the boost condition. The associated local Legendre transform is
\begin{align}
p_H={}&Ts+\mu n+v_iT^{0i} +\sum_{X,a}\rho_X^a
       \frac{\partial T^{00}}{\partial\rho_X^a}-T^{00}.\label{p:pH}
\end{align}
We distinguish $p_H$ from the Lorentz-scalar pressure until the covariant reconstruction. Integrating the gradient variations by parts gives
\begin{align}
\frac{\delta H}{\delta\Sigma}
 &=\frac{\partial T^{00}}{\partial\Sigma}
   -\partial_i\frac{\partial T^{00}}{\partial(\partial_i\Sigma)},\notag\\
\frac{\delta H}{\delta\Sigma^\dagger}
 &=\frac{\partial T^{00}}{\partial\Sigma^\dagger}
   -\partial_i\frac{\partial T^{00}}{\partial(\partial_i\Sigma^\dagger)}.\label{p:functional}
\end{align}
In the differential of \eqref{p:pH}, the variations of $s,n,T^{0i},\rho_L^a,$ and $\rho_R^a$ cancel against the corresponding terms in $dT^{00}$. The remaining terms give
\begin{align}
dp_H={}&s\,dT+n\,d\mu+T^{0i}dv_i +\sum_{X,a}\rho_X^a d\!\left(\frac{\partial T^{00}}{\partial\rho_X^a}\right)
\notag\\&-\operatorname{tr}\!\left[\frac{\partial T^{00}}{\partial\Sigma} d\Sigma+\frac{\partial T^{00}}{\partial\Sigma^\dagger} d\Sigma^\dagger\right]
\notag\\&-\operatorname{tr}\!\left[\frac{\partial T^{00}}{\partial(\partial_{i}\Sigma)}d(\partial_i\Sigma)+\frac{\partial T^{00}}{\partial(\partial_{i}\Sigma^\dagger)}d(\partial_i\Sigma^\dagger)\right].\label{p:GD}
\end{align}
This identity follows from the Legendre transform and does not require $T^{00}$ to be homogeneous in its arguments.

Using the momentum-density bracket gives
\begin{align}
\partial_0T^{0i}={}&-s\partial_iT-n\partial_i\mu-T^{0j}\partial_i v_j
                   -\partial_j(v_jT^{0i})\notag\\
 &-\sum_{X,a}\rho_X^a\partial_i\!\left(\frac{\partial T^{00}}{\partial\rho_X^a}\right)\notag\\
 &+\operatorname{tr}\!\left[\frac{\delta H}{\delta\Sigma}\partial_i\Sigma+\frac{\delta H}{\delta\Sigma^\dagger}\partial_i\Sigma^\dagger\right].\label{p:momraw}
\end{align}
The scalar-density term, for example, is
\begin{equation*}
-\int d^3y\,T(\mathbf y)s(\mathbf x)\partial_{x^i}\delta(\mathbf x-\mathbf y)=-s\partial_iT.
\end{equation*}
Taking a spatial derivative in \eqref{p:GD} and using
\begin{align}
&\operatorname{tr}\!\left[\frac{\partial T^{00}}{\partial\Sigma}\partial_i\Sigma+\frac{\partial T^{00}}{\partial(\partial_{j}\Sigma)}\partial_i\partial_j\Sigma\right]\notag\\
&\quad=\operatorname{tr}\!\left[\frac{\delta H}{\delta\Sigma}\partial_i\Sigma\right]
    +\partial_j\operatorname{tr}\!\left[\frac{\partial T^{00}}{\partial(\partial_{j}\Sigma)}\partial_i\Sigma\right]
\end{align}
together with the corresponding identity for $\Sigma^\dagger$ converts \eqref{p:momraw} to $\partial_0T^{0i}=-\partial_jT^{ji}$, where
\begin{align}
T^{ji}&=v_jT^{0i}+p_H\delta_{ij}\notag\\
 &\quad+\operatorname{tr}\!\left[\frac{\partial T^{00}}{\partial(\partial_{j}\Sigma)}\partial_i\Sigma+\frac{\partial T^{00}}{\partial(\partial_{j}\Sigma^\dagger)}\partial_i\Sigma^\dagger\right].\label{p:Tji}
\end{align}
This expression follows from the momentum bracket alone. Spatial rotations supply the condition that makes it symmetric.

\subsection{Spatial rotations}
At corresponding points under a rotation, $x'^i=x^i+\theta_{ij}x^j$ with $\theta_{ij}=-\theta_{ji}$, the components vary as $\delta T^{0i}=\theta_{ij}T^{0j}$ and $\delta(\partial_i\Sigma)=\theta_{ij}\partial_j\Sigma$. The same rule applies to $\partial_i\Sigma^\dagger$. Rotational invariance of the Hamiltonian therefore gives
\begin{align}
&0=\delta T^{00}
 =\theta_{ij}\Bigl[v_iT^{0j}+\operatorname{tr}\!\Bigl(
 \frac{\partial T^{00}}{\partial(\partial_i\Sigma)}\partial_j\Sigma\notag\\
&\qquad+\frac{\partial T^{00}}{\partial(\partial_i\Sigma^\dagger)}\partial_j\Sigma^\dagger\Bigr)\Bigr]\notag\\
&\quad=\tfrac12\theta_{ij}(T^{ij}-T^{ji}).\label{p:rotation}
\end{align}
The arbitrariness of the antisymmetric $\theta_{ij}$ implies $T^{ij}=T^{ji}$. At fixed coordinates, each field variation also includes the orbital term $-\theta_{k\ell}x^\ell\partial_k$. These terms reproduce the coordinate variation of $T^{00}$ and leave the same symmetry condition.

\subsection{Local energy current}
The energy current can be obtained directly from the Hamilton equations. Expanding the time derivative of the Hamiltonian density gives
\begin{align}
\partial_0T^{00}={}&T\partial_0s+\mu\partial_0n+v_i\partial_0T^{0i}\notag\\
 &+\sum_{X,a}\frac{\partial T^{00}}{\partial\rho_X^a}\partial_0\rho_X^a\notag\\
 &+\operatorname{tr}\!\left[\frac{\partial T^{00}}{\partial\Sigma}\partial_0\Sigma+\frac{\partial T^{00}}{\partial\Sigma^\dagger}\partial_0\Sigma^\dagger\right]\notag\\
 &+\operatorname{tr}\!\left[\frac{\partial T^{00}}{\partial(\partial_{i}\Sigma)}\partial_i\partial_0\Sigma
                  +\frac{\partial T^{00}}{\partial(\partial_{i}\Sigma^\dagger)}\partial_i\partial_0\Sigma^\dagger\right].\label{p:chain}
\end{align}
For the $\Sigma$ contribution,
\begin{align}
\operatorname{tr}\!\left[\frac{\partial T^{00}}{\partial(\partial_{i}\Sigma)}\partial_i\partial_0\Sigma\right]
 &=\partial_i\operatorname{tr}\!\left[\frac{\partial T^{00}}{\partial(\partial_{i}\Sigma)}\partial_0\Sigma\right]\notag\\
 &\quad-\operatorname{tr}\!\left[\left(\partial_i\frac{\partial T^{00}}{\partial(\partial_{i}\Sigma)}\right)\partial_0\Sigma\right].
\end{align}
The total derivative contributes to the energy flux. Its remainder combines with the ordinary matrix derivative to give the functional derivative in \eqref{p:functional}.

The internal charge and order-parameter dynamics cancel in the energy balance. For either chirality,
\begin{equation*}
f^{abc}\frac{\partial T^{00}}{\partial\rho_X^a}
       \frac{\partial T^{00}}{\partial\rho_X^b}\rho_X^c=0,
\end{equation*}
by antisymmetry of $f^{abc}$ in $a,b$. The charge--order-parameter couplings cancel against the corresponding terms in the $\Sigma$ and $\Sigma^\dagger$ equations, as shown separately in Appendix~\ref{app:internal}. Only their advective contributions remain:
\begin{align}
&\sum_{X,a}\frac{\partial T^{00}}{\partial\rho_X^a}\partial_0\rho_X^a +\operatorname{tr}\!\left[\frac{\delta H}{\delta\Sigma}\partial_0\Sigma
                 +\frac{\delta H}{\delta\Sigma^\dagger}\partial_0\Sigma^\dagger\right]
\notag\\=&-\sum_{X,a}\frac{\partial T^{00}}{\partial\rho_X^a}\partial_i(v_i\rho_X^a) -v_i\operatorname{tr}\!\left[\frac{\delta H}{\delta\Sigma}\partial_i\Sigma
                 +\frac{\delta H}{\delta\Sigma^\dagger}\partial_i\Sigma^\dagger\right].
\end{align}
The last term cancels the order-parameter force in $v_i\partial_0T^{0i}$. With $\partial_0s=-\partial_i(v_is)$ and $\partial_0n=-\partial_i(v_in)$, the energy balance becomes
\begin{align}
\partial_0T^{00}={}&-\partial_i(v_iTs)-\partial_i(v_i\mu n) -\partial_i\!\left[v_i\sum_{X,a}\rho_X^a
                         \frac{\partial T^{00}}{\partial\rho_X^a}\right]
\notag\\&-v_iT^{0j}\partial_i v_j-v_i\partial_j(v_jT^{0i})
\notag\\&+\partial_i\operatorname{tr}\!\left[\frac{\partial T^{00}}{\partial(\partial_{i}\Sigma)}\partial_0\Sigma+\frac{\partial T^{00}}{\partial(\partial_{i}\Sigma^\dagger)}\partial_0\Sigma^\dagger\right].
\end{align}
Relabeling $i,j$ in the first of the two momentum terms combines them into $-\partial_i(v_i v_jT^{0j})$. Hence
\begin{align}
\partial_0T^{00}&=-\partial_iT^{i0},\notag\\
T^{i0}&=v_i(T^{00}+p_H)\notag\\
&\quad-\operatorname{tr}\!\left[\frac{\partial T^{00}}{\partial(\partial_{i}\Sigma)}\partial_0\Sigma+\frac{\partial T^{00}}{\partial(\partial_{i}\Sigma^\dagger)}\partial_0\Sigma^\dagger\right].\label{p:Ti0}
\end{align}
Local energy conservation has now fixed the flux. Relating this flux to momentum density is the additional step required by boosts.

\subsection{Boost condition}
For the orbital boost current in the spinless case considered here,
\begin{equation}
\partial_\mu(x^iT^{\mu0}-x^0T^{\mu i})=T^{i0}-T^{0i},
\end{equation}
after the already derived energy and momentum equations are applied. Local conservation of this boost current imposes
\begin{equation}\label{p:boost}T^{i0}=T^{0i}.\end{equation}
Thus boost invariance constrains the Hamiltonian through a relation not required by energy conservation alone. Appendix~\ref{app:bracket} obtains the same flux from the smeared energy-density bracket and shows how \eqref{p:boost} gives the classical Dirac--Schwinger relation.

\section{Ordinary-fluid limit}
\subsection{Thermodynamic Legendre transform}
The ordinary-fluid limit supplies the velocity derivatives used in the chiral calculation. Set the chiral charge densities and pion gradients to zero, so that $T=\partial\epsilon/\partial s$ and $\mu=\partial\epsilon/\partial n$. Denote the momentum conjugate by $v_{0i}=\partial\epsilon/\partial T^{0i}$; in this section we write $v_i=v_{0i}$. The boost condition is then
\begin{align}
T^{0i}&=(\epsilon+p_0)v_i,\notag\\
p_0&=s\frac{\partial\epsilon}{\partial s}
 +n\frac{\partial\epsilon}{\partial n}
 +T^{0j}\frac{\partial\epsilon}{\partial T^{0j}}-\epsilon.\label{p:fluid}
\end{align}
Identifying $p_0$ as the rest-frame pressure also uses local equilibrium and extensivity of the homogeneous rest energy. At rest, the first law and extensivity give $\epsilon|_{\mathbf v=0}+p_0=Ts+\mu n$. Its differential gives the Gibbs--Duhem relation~\cite{LandauLifshitz,AnderssonComer2021}. This assumption concerns the homogeneous rest energy, not the full Hamiltonian with gradients.

\subsection{Velocity derivatives at fixed proper densities}
The Hessian of $\epsilon$ is symmetric. Differentiating \eqref{p:fluid} gives
\begin{align}
\frac{\partial(\epsilon+p_0)}{\partial T^{0\ell}}
 &=v_\ell+s\frac{\partial v_\ell}{\partial s}+n\frac{\partial v_\ell}{\partial n}
             +T^{0j}\frac{\partial v_\ell}{\partial T^{0j}},\notag\\
\delta_{i\ell}
 &=(\epsilon+p_0)\frac{\partial v_\ell}{\partial T^{0i}}+v_i v_\ell\notag\\
 &\quad+v_i\left(s\frac{\partial}{\partial s}+n\frac{\partial}{\partial n}
             +T^{0j}\frac{\partial}{\partial T^{0j}}\right)v_\ell.
\end{align}
The differential operator appearing in this identity is
\begin{align}
\mathcal{D}_i={}&(\epsilon+p_0)\frac{\partial}{\partial T^{0i}}\notag\\
 &+v_i\left(T^{0j}\frac{\partial}{\partial T^{0j}}
           +s\frac{\partial}{\partial s}+n\frac{\partial}{\partial n}\right).\label{p:D}
\end{align}
It satisfies
\begin{equation}\label{p:Dv}
\mathcal{D}_i v_j=\delta_{ij}-v_iv_j,\quad \mathcal{D}_i s=v_i s,\quad \mathcal{D}_i n=v_i n.
\end{equation}
For $\mathbf v^2<1$, define the Lorentz factor
\begin{equation}\label{p:gamma}
\gamma=(1-\mathbf v^2)^{-1/2},\qquad \mathcal{D}_i\gamma=v_i\gamma.
\end{equation}
The second identity follows directly from \eqref{p:Dv}. Together with $\mathcal{D}_i s=v_i s$ and $\mathcal{D}_i n=v_i n$, it gives
\begin{equation}
\mathcal{D}_i(s/\gamma)=0,\qquad \mathcal{D}_i(n/\gamma)=0.
\end{equation}
For $\mathbf v^2<1$, the matrix $\delta_{ij}-v_iv_j$ has inverse $\delta_{ij}+v_iv_j/(1-\mathbf v^2)$ and is not a projector. We assume that the ordinary-fluid energy density is finite and that the thermodynamic relations between $(s,n,T^{0i})$ and $(\mathbf v,s/\gamma,n/\gamma)$ can be solved in either direction, with finite derivatives near the chosen state. The chain rule then expresses $\mathcal D_i$ as
\begin{equation}\label{p:Dvelocity}
\mathcal{D}_i=(\delta_{i\ell}-v_iv_\ell)
 \left.\frac{\partial}{\partial v_\ell}\right|_{s/\gamma,n/\gamma}.
\end{equation}
All velocity derivatives below are ordinary derivatives on state space, taken at the indicated fixed proper densities.

The same identities determine the ordinary-fluid velocity dependence:
\begin{align}
\mathcal{D}_i\epsilon&=2v_i(\epsilon+p_0),\notag\\
\mathcal{D}_iT^{0j}&=(\epsilon+p_0)\delta_{ij}+v_iT^{0j}\notag\\
 &=v_j\mathcal{D}_i(\epsilon+p_0)+(\epsilon+p_0)(\delta_{ij}-v_iv_j).
\end{align}
For nonzero $\mathbf v$, comparison yields $\mathcal{D}_i(\epsilon+p_0)=2v_i(\epsilon+p_0)$. Since the thermodynamic quantities and their derivatives approach their rest-frame values as the velocity tends to zero, the same identity holds at rest. Thus $p_0$ and $(\epsilon+p_0)/\gamma^2$ depend only on $s/\gamma,n/\gamma$, and
\begin{align}
\epsilon&=\gamma^2(\epsilon|_{\mathbf v=0}+p_0)-p_0,\notag\\
T^{0i}&=\gamma^2(\epsilon|_{\mathbf v=0}+p_0)v_i.\label{p:fluidresult}
\end{align}
Here $\epsilon|_{\mathbf v=0}$ is evaluated at the proper densities $s/\gamma,n/\gamma$. Differentiating $\epsilon-v_iT^{0i}=\epsilon|_{\mathbf v=0}$ with these arguments shows that the proper temperature and chemical potential are $\gamma T$ and $\gamma\mu$.

\section{Boost constraints in the chiral sector}
\subsection{Expansion at fixed equal-time variables}
We now restore the chiral contribution $\Delta\epsilon=T^{00}-\epsilon=O[(\rho_{L,R},\partial\Sigma)^2]$. Holding the independent fluid variables fixed, the full conjugates become
\begin{equation}\label{p:split}
\begin{aligned}
T&=\frac{\partial\epsilon}{\partial s}+\frac{\partial\Delta\epsilon}{\partial s},
\qquad\mu=\frac{\partial\epsilon}{\partial n}+\frac{\partial\Delta\epsilon}{\partial n},\\
v_i&=v_{0i}+\frac{\partial\Delta\epsilon}{\partial T^{0i}}.
\end{aligned}
\end{equation}
Each derivative of $\Delta\epsilon$ is quadratic in the chiral charge densities and pion gradients. The Legendre transforms give
\begin{equation}
\begin{aligned}
p_H-p_0={}&s\frac{\partial\Delta\epsilon}{\partial s}
 +n\frac{\partial\Delta\epsilon}{\partial n}
 +T^{0j}\frac{\partial\Delta\epsilon}{\partial T^{0j}}\\
 &+\sum_{X,a}\rho_X^a\frac{\partial\Delta\epsilon}{\partial\rho_X^a}-\Delta\epsilon.
\end{aligned}
\end{equation}
In $T^{i0}-T^{0i}$, the ordinary-fluid contribution vanishes by \eqref{p:fluid}. The product
\begin{equation}
\begin{aligned}
&\frac{\partial\Delta\epsilon}{\partial T^{0i}}
 \biggl(s\frac{\partial\Delta\epsilon}{\partial s}
       +n\frac{\partial\Delta\epsilon}{\partial n}
       +T^{0j}\frac{\partial\Delta\epsilon}{\partial T^{0j}}
       +\sum_{X,a}\rho_X^a\frac{\partial\Delta\epsilon}{\partial\rho_X^a}\biggr)\\
&\qquad=O[(\rho_{L,R},\partial\Sigma)^4]
\end{aligned}
\end{equation}
is therefore discarded. Using \eqref{p:D}, the remaining terms are
\begin{align}
&T^{i0}-T^{0i}=\mathcal{D}_i\Delta\epsilon
 +v_{0i}\sum_{X,a}\rho_X^a\frac{\partial T^{00}}{\partial\rho_X^a}\notag\\
 &\quad-\operatorname{tr}\!\left[\frac{\partial T^{00}}{\partial(\partial_i\Sigma)}\partial_0\Sigma
              +\frac{\partial T^{00}}{\partial(\partial_i\Sigma^\dagger)}\partial_0\Sigma^\dagger\right]
 +O[(\rho_{L,R},\partial\Sigma)^4].\label{p:residualstart}
\end{align}

In the quadratic terms of this section and Sec.~\ref{sec:boostsolution}, we write $v_i=v_{0i}$. Velocity derivatives hold $s/\gamma,n/\gamma$, the chiral densities, $\Sigma$, and its spatial derivatives fixed. Replacing $v_{0i}$ by the full velocity in the leading $\Sigma$ equation changes $\partial_0\Sigma$ only at cubic order, and its contribution to the flux only at quartic order.

The first contribution is obtained by differentiating only the coefficient functions:

\begin{align}
&\mathcal{D}_i\Delta\epsilon
 =(\delta_{i\ell}-v_i v_\ell)\Bigg[\frac{\partial f_{jk}}{\partial v_\ell}\operatorname{tr}(\partial_j\Sigma\partial_k\Sigma^\dagger)\notag\\
 &\quad+\frac12\frac{\partial a}{\partial v_\ell}\operatorname{tr}\bigl(\rho_L-\Sigma\rho_R\Sigma^\dagger\bigr)^2
 +\frac12\frac{\partial a_v}{\partial v_\ell}\operatorname{tr}\bigl(\rho_L+\Sigma\rho_R\Sigma^\dagger\bigr)^2\notag\\
 &\quad-i\frac{\partial c_j}{\partial v_\ell}
 \operatorname{tr}\!\left[\bigl(\rho_L-\Sigma\rho_R\Sigma^\dagger\bigr)\Sigma\partial_j\Sigma^\dagger\right]\Bigg].\label{p:velocitypart}
\end{align}

\vspace*{0pt}
\subsection{Density derivatives}
The density contribution to \eqref{p:residualstart} requires derivatives with respect to both chiralities. Differentiating the two factors in each square and using cyclicity gives
\begin{widetext}
\begin{align}
\sum_a\rho_L^a\frac{\partial}{\partial\rho_L^a}\left(\frac a2\operatorname{tr}\bigl(\rho_L-\Sigma\rho_R\Sigma^\dagger\bigr)^2\right) &=a\operatorname{tr}\!\left[\rho_L\bigl(\rho_L-\Sigma\rho_R\Sigma^\dagger\bigr)\right],\notag\\
\sum_a\rho_R^a\frac{\partial}{\partial\rho_R^a}\left(\frac a2\operatorname{tr}\bigl(\rho_L-\Sigma\rho_R\Sigma^\dagger\bigr)^2\right) &=-a\operatorname{tr}\!\left[\Sigma\rho_R\Sigma^\dagger\bigl(\rho_L-\Sigma\rho_R\Sigma^\dagger\bigr)\right],\notag\\
\sum_a\rho_L^a\frac{\partial}{\partial\rho_L^a}\left(\frac{a_v}{2}\operatorname{tr}\bigl(\rho_L+\Sigma\rho_R\Sigma^\dagger\bigr)^2\right) &=a_v\operatorname{tr}\!\left[\rho_L\bigl(\rho_L+\Sigma\rho_R\Sigma^\dagger\bigr)\right],\notag\\
\sum_a\rho_R^a\frac{\partial}{\partial\rho_R^a}\left(\frac{a_v}{2}\operatorname{tr}\bigl(\rho_L+\Sigma\rho_R\Sigma^\dagger\bigr)^2\right) &=a_v\operatorname{tr}\!\left[\Sigma\rho_R\Sigma^\dagger\bigl(\rho_L+\Sigma\rho_R\Sigma^\dagger\bigr)\right].\label{p:rhosquares}
\end{align}

For the density--gradient term in its original form,

\begin{align}
\sum_a\rho_L^a\frac{\partial}{\partial\rho_L^a}
 \left[-ic_j\operatorname{tr}(\rho_L\Sigma\partial_j\Sigma^\dagger+\Sigma^\dagger\partial_j\Sigma\rho_R)\right]
 &=-ic_j\operatorname{tr}(\rho_L\Sigma\partial_j\Sigma^\dagger),\notag\\
\sum_a\rho_R^a\frac{\partial}{\partial\rho_R^a}
 \left[-ic_j\operatorname{tr}(\rho_L\Sigma\partial_j\Sigma^\dagger+\Sigma^\dagger\partial_j\Sigma\rho_R)\right]
 &=-ic_j\operatorname{tr}(\Sigma^\dagger\partial_j\Sigma\rho_R) =+ic_j\operatorname{tr}(\Sigma\rho_R\Sigma^\dagger\Sigma\partial_j\Sigma^\dagger).\label{p:rhocross}
\end{align}
\end{widetext}
Adding these derivatives yields

\begin{align}
&\sum_{X,a}\rho_X^a\frac{\partial T^{00}}{\partial\rho_X^a}
 =a\operatorname{tr}\bigl(\rho_L-\Sigma\rho_R\Sigma^\dagger\bigr)^2\notag\\
&\quad+a_v\operatorname{tr}\bigl(\rho_L+\Sigma\rho_R\Sigma^\dagger\bigr)^2\notag\\
&\quad-ic_j\operatorname{tr}\!\left[\bigl(\rho_L-\Sigma\rho_R\Sigma^\dagger\bigr)\Sigma\partial_j\Sigma^\dagger\right].\label{p:rhosum}
\end{align}
The pure-gradient term contributes no density derivative. As a check, a common rescaling of the charge densities multiplies each density square by degree two and each mixed term by degree one, exactly as in \eqref{p:rhosum}. Multiplication by $v_i$ gives the second contribution to \eqref{p:residualstart}.

\subsection{Order-parameter contribution to the energy flux}
Varying the gradient arguments before using unitarity gives
\begin{align}
\delta T^{00}\big|_{\partial\Sigma,\partial\Sigma^\dagger}
 ={}&\operatorname{tr}\!\left[(f_{ij}\partial_j\Sigma^\dagger-ic_i\rho_R\Sigma^\dagger)\delta(\partial_i\Sigma)\right]\notag\\
 &+\operatorname{tr}\!\left[(f_{ij}\partial_j\Sigma-ic_i\rho_L\Sigma)\delta(\partial_i\Sigma^\dagger)\right].
\end{align}
Each variation has been cyclically moved to the right. Consequently,
\begin{align}
\frac{\partial T^{00}}{\partial(\partial_{i}\Sigma)}&=f_{ij}\partial_j\Sigma^\dagger-ic_i\rho_R\Sigma^\dagger,\notag\\
\frac{\partial T^{00}}{\partial(\partial_{i}\Sigma^\dagger)}&=f_{ij}\partial_j\Sigma-ic_i\rho_L\Sigma.\label{p:gradpartials}
\end{align}
After taking the traces, the two $f_{ij}$ contributions are complex conjugates. The mixed contributions involve different charge matrices and must be retained separately. If instead the gradient variation is constrained at fixed $\Sigma$, one finds
\begin{align}
&-ic_i\operatorname{tr}\!\left[\rho_L\Sigma\delta(\partial_i\Sigma^\dagger)
       +\Sigma^\dagger\delta(\partial_i\Sigma)\rho_R\right]\notag\\
&\quad=ic_i\operatorname{tr}\!\left[\Sigma^\dagger\bigl(\rho_L-\Sigma\rho_R\Sigma^\dagger\bigr)\delta(\partial_i\Sigma)\right],
\end{align}
where $\delta(\partial_i\Sigma^\dagger)=-\Sigma^\dagger\delta(\partial_i\Sigma)\Sigma^\dagger$. This constrained variation already contains both mixed terms. Adding its complex conjugate would count them twice. Rewriting a real variation with $2\operatorname{Re}$ therefore changes neither ordinary derivative in \eqref{p:gradpartials} nor its normalization.

Generator completeness gives the matrix forms of the conjugates to $\rho_L^a$ and $\rho_R^a$:
\begin{align}
\sum_a\frac{\partial T^{00}}{\partial\rho_L^a}\frac{\lambda^a}{2}
 &=\frac a2\bigl(\rho_L-\Sigma\rho_R\Sigma^\dagger\bigr)
 +\frac{a_v}{2}\bigl(\rho_L+\Sigma\rho_R\Sigma^\dagger\bigr)\notag\\
 &\quad-\frac{ic_j}{2}\Sigma\partial_j\Sigma^\dagger,\notag\\[3pt]
\sum_a\frac{\partial T^{00}}{\partial\rho_R^a}\frac{\lambda^a}{2}
 &=-\frac a2\Sigma^\dagger\bigl(\rho_L-\Sigma\rho_R\Sigma^\dagger\bigr)\Sigma\notag\\
 &\quad+\frac{a_v}{2}\Sigma^\dagger\bigl(\rho_L+\Sigma\rho_R\Sigma^\dagger\bigr)\Sigma
 -\frac{ic_j}{2}\Sigma^\dagger\partial_j\Sigma.\label{p:rhoconj}
\end{align}

The factors $1/2$ here follow from $\sum_a(\lambda^a/2)\operatorname{tr}[(\lambda^a/2)Z]=Z/2$ for a traceless matrix $Z$. Inserting \eqref{p:rhoconj} into the Hamilton equation gives the explicit cancellation

\begin{widetext}
\begin{align}
\partial_0\Sigma
 &=-v_j\partial_j\Sigma
 -i\left[\frac a2\bigl(\rho_L-\Sigma\rho_R\Sigma^\dagger\bigr)+\frac{a_v}{2}\bigl(\rho_L+\Sigma\rho_R\Sigma^\dagger\bigr)-\frac{ic_j}{2}\Sigma\partial_j\Sigma^\dagger\right]\Sigma\notag\\
&\quad+i\Sigma\left[-\frac a2\Sigma^\dagger\bigl(\rho_L-\Sigma\rho_R\Sigma^\dagger\bigr)\Sigma
 +\frac{a_v}{2}\Sigma^\dagger\bigl(\rho_L+\Sigma\rho_R\Sigma^\dagger\bigr)\Sigma-\frac{ic_j}{2}\Sigma^\dagger\partial_j\Sigma\right]\notag\\
&=-v_j\partial_j\Sigma-\frac{ia}{2}\bigl(\rho_L-\Sigma\rho_R\Sigma^\dagger\bigr)\Sigma-\frac{ia}{2}\bigl(\rho_L-\Sigma\rho_R\Sigma^\dagger\bigr)\Sigma\notag\\
&\quad-\frac{ia_v}{2}\bigl(\rho_L+\Sigma\rho_R\Sigma^\dagger\bigr)\Sigma+\frac{ia_v}{2}\bigl(\rho_L+\Sigma\rho_R\Sigma^\dagger\bigr)\Sigma
 -\frac{c_j}{2}\Sigma\partial_j\Sigma^\dagger\Sigma+\frac{c_j}{2}\partial_j\Sigma\notag\\
&=-(v_j-c_j)\partial_j\Sigma-ia\bigl(\rho_L-\Sigma\rho_R\Sigma^\dagger\bigr)\Sigma.\label{p:Sigma}
\end{align}

Its conjugate is $\partial_0\Sigma^\dagger=-(v_j-c_j)\partial_j\Sigma^\dagger+ia\Sigma^\dagger\bigl(\rho_L-\Sigma\rho_R\Sigma^\dagger\bigr)$. The terms proportional to $a_v$ cancel, while the two mixed contributions add because $\Sigma\partial_j\Sigma^\dagger\Sigma=-\partial_j\Sigma$.

The last contribution to $T^{i0}-T^{0i}$ contains four products:

\begin{align}
&-\operatorname{tr}\!\left[\frac{\partial T^{00}}{\partial(\partial_{i}\Sigma)}\partial_0\Sigma+\frac{\partial T^{00}}{\partial(\partial_{i}\Sigma^\dagger)}\partial_0\Sigma^\dagger\right]\notag\\
&\quad=-\operatorname{tr}\!\Big[(f_{ij}\partial_j\Sigma^\dagger-ic_i\rho_R\Sigma^\dagger)
              (-(v_k-c_k)\partial_k\Sigma-ia\bigl(\rho_L-\Sigma\rho_R\Sigma^\dagger\bigr)\Sigma)\notag\\
&\quad+(f_{ij}\partial_j\Sigma-ic_i\rho_L\Sigma)
              (-(v_k-c_k)\partial_k\Sigma^\dagger+ia\Sigma^\dagger\bigl(\rho_L-\Sigma\rho_R\Sigma^\dagger\bigr))\Big]\notag\\
&\quad=f_{ij}(v_k-c_k)\operatorname{tr}(\partial_j\Sigma^\dagger\partial_k\Sigma
                                     +\partial_j\Sigma\partial_k\Sigma^\dagger)\notag\\
&\qquad+iaf_{ij}\operatorname{tr}[\partial_j\Sigma^\dagger\bigl(\rho_L-\Sigma\rho_R\Sigma^\dagger\bigr)\Sigma
                               -\partial_j\Sigma\Sigma^\dagger\bigl(\rho_L-\Sigma\rho_R\Sigma^\dagger\bigr)]\notag\\
&\qquad-ic_i(v_j-c_j)\operatorname{tr}(\rho_R\Sigma^\dagger\partial_j\Sigma
                                    +\rho_L\Sigma\partial_j\Sigma^\dagger)\notag\\
&\qquad+ac_i\operatorname{tr}[\rho_R\Sigma^\dagger\bigl(\rho_L-\Sigma\rho_R\Sigma^\dagger\bigr)\Sigma-\rho_L\bigl(\rho_L-\Sigma\rho_R\Sigma^\dagger\bigr)]\notag\\
&\quad=\bigl[(v_j-c_j)f_{ik}+(v_k-c_k)f_{ij}\bigr]\operatorname{tr}(\partial_j\Sigma\partial_k\Sigma^\dagger)\notag\\
&\qquad+i[2af_{ij}-c_i(v_j-c_j)]\operatorname{tr}[\bigl(\rho_L-\Sigma\rho_R\Sigma^\dagger\bigr)\Sigma\partial_j\Sigma^\dagger]
 -ac_i\operatorname{tr}\bigl(\rho_L-\Sigma\rho_R\Sigma^\dagger\bigr)^2.\label{p:products}
\end{align}
\end{widetext}
The four products simplify for distinct reasons. The gradient traces coincide by \eqref{p:gradreal}. The gradient--density terms add by cyclicity and $\partial_j\Sigma\Sigma^\dagger=-\Sigma\partial_j\Sigma^\dagger$, while the density--gradient term reproduces the original mixed invariant. In the density product, moving $\Sigma$ around the trace leaves $(\Sigma\rho_R\Sigma^\dagger-\rho_L)\bigl(\rho_L-\Sigma\rho_R\Sigma^\dagger\bigr)$, fixing its negative sign. Symmetry in $j,k$ then gives the displayed gradient coefficient. Each step retains the two independent gradient derivatives.

\subsection{Independent local structures}
Adding \eqref{p:velocitypart}, \eqref{p:rhosum}, and \eqref{p:products} gives the coefficients of the four local structures:
\begin{widetext}
\begin{align}
T^{i0}-T^{0i} &=\left[\tfrac12\mathcal{D}_i a+v_i a-ac_i\right]\operatorname{tr}\bigl(\rho_L-\Sigma\rho_R\Sigma^\dagger\bigr)^2
 +\left[\tfrac12\mathcal{D}_i a_v+v_i a_v\right]\operatorname{tr}\bigl(\rho_L+\Sigma\rho_R\Sigma^\dagger\bigr)^2\notag\\
&\quad+i[-\mathcal{D}_i c_j-v_i c_j+2af_{ij}-c_i v_j+c_i c_j]
                  \operatorname{tr}[\bigl(\rho_L-\Sigma\rho_R\Sigma^\dagger\bigr)\Sigma\partial_j\Sigma^\dagger]\notag\\
&\quad+[\mathcal{D}_i f_{jk}+(v_j-c_j)f_{ik}+(v_k-c_k)f_{ij}]
                  \operatorname{tr}(\partial_j\Sigma\partial_k\Sigma^\dagger)+O[(\rho_{L,R},\partial\Sigma)^4].\label{p:fourstructures}
\end{align}
\end{widetext}
The four coefficients can be isolated by choosing the local chiral data at a fixed fluid state and fixed $\Sigma$. With zero gradients and $\rho_R=-\Sigma^\dagger\rho_L\Sigma$, only the first square survives for nonzero Hermitian traceless $\rho_L$. Its coefficient must vanish. Choosing instead $\rho_R=+\Sigma^\dagger\rho_L\Sigma$ isolates the second square.

Next set both densities to zero and choose local gradients $\partial_j\Sigma=i k_j(\lambda^1/2)\Sigma$, which can be realized by exponentiating a chiral generator with a phase linear in position. The gradient trace is $k_jk_k/2$. Its coefficient is symmetric in $j,k$ and has zero contraction for every choice of real $k_j$, so each component vanishes. Finally, keep the same gradient data, take a nonzero Hermitian $\rho_L$ proportional to $\lambda^1$, and set $\Sigma\rho_R\Sigma^\dagger=-\rho_L$. Then
\begin{equation*}
i\operatorname{tr}[\bigl(\rho_L-\Sigma\rho_R\Sigma^\dagger\bigr)\Sigma\partial_j\Sigma^\dagger]
 =k_j\operatorname{tr}(\rho_L\lambda^1).
\end{equation*}
The trace is nonzero, so varying $k_j$ sets the mixed coefficient to zero. The density and gradient coefficients have already vanished, which is why this last choice isolates the remaining structure.

\subsection{Differential equations}
\eqref{p:boost} therefore gives
\begin{align}
\mathcal D_i a&=2a(c_i-v_i),\qquad \mathcal D_i a_v=-2v_i a_v,\notag\\
\mathcal D_i c_j&=2af_{ij}+c_i c_j-c_i v_j-v_i c_j,\notag\\
\mathcal D_i f_{jk}&=(c_j-v_j)f_{ik}+(c_k-v_k)f_{ij}.\label{p:fourPDE}
\end{align}
Here $\mathcal{D}_i=(\delta_{i\ell}-v_i v_\ell)\partial/\partial v_\ell$ at fixed $s/\gamma,n/\gamma$. These equations follow entirely from the equal-time brackets and Hamiltonian; their solution will determine the covariant structure.

\section{Solution of the boost equations}\label{sec:boostsolution}
To integrate the boost equations, specify the Hamiltonian at rest. Its three functions $f_s^2,f_t^2,f_v^2$ depend on $s/\gamma,n/\gamma$, which are held fixed throughout this section. With the ordinary-fluid variables fixed, $f_s$ controls the spatial pion-gradient term, $f_t$ the axial-density term and temporal response, and $f_v$ the vector-density term. The initial data from (4) of Ref.~\cite{Son2000} are
\begin{align}
a(0)&=\frac2{f_t^2},&a_v(0)&=\frac2{f_v^2},\notag\\
c_i(0)&=0,&f_{ij}(0)&=\frac{f_s^2}{4}\delta_{ij}.\label{p:restdata}
\end{align}
Indeed its axial term is $f_t^{-2}\operatorname{tr}\bigl(\rho_L-\Sigma\rho_R\Sigma^\dagger\bigr)^2$, and its equation for $\Sigma$ is $\partial_0\Sigma=-2i\bigl(\rho_L-\Sigma\rho_R\Sigma^\dagger\bigr)\Sigma/f_t^2$. Both the published version of Ref.~\cite{Son2000} (p.~3773, below (13)) and its arXiv version 2 print $a=2f_s^{-2}$. This is inconsistent with the axial term in (4) and the evolution equation (5) of that paper when $f_s\ne f_t$. The rest-frame Hamiltonian fixes $a(0)=2/f_t^2$, as used here.

The vector equation gives $\mathcal{D}_i(\gamma^2a_v)=0$, since $\mathcal{D}_i\gamma=v_i\gamma$. Its rest value yields $a_v=2/(\gamma^2 f_v^2)$.

Rotational symmetry reduces the coupled equations to a radial problem. Choose $\mathbf v=(0,0,v)$, so $c_1=c_2=0$, $f_{12}=f_{13}=f_{23}=0$, and $f_{11}=f_{22}$. The $i=3,j=k=1,2$ equations keep the transverse coefficients fixed at $f_s^2/4$. In terms of the rapidity $\eta=\operatorname{artanh}v$, for which $d/d\eta=(1-v^2)d/dv$, the remaining equations read
\begin{align}
\frac{da}{d\eta}&=2a(c_3-v),\notag\\
\frac{dc_3}{d\eta}&=2af_{33}+c_3^2-2vc_3,\notag\\
\frac{df_{33}}{d\eta}&=2(c_3-v)f_{33}.\label{p:radial}
\end{align}
The nonlinear terms simplify in the following combinations:
\begin{align}
&\frac{d}{d\eta}\left(\frac2a\right)=\frac{4(v-c_3)}a,\notag\\
&\frac{d}{d\eta}\left(\frac{2(v-c_3)}a\right)
 =\frac2a+\frac{2(v-c_3)^2}a-4f_{33},\notag\\
&\frac{d}{d\eta}\left(\frac{2(v-c_3)^2}a-4f_{33}\right)
 =\frac{4(v-c_3)}a.\label{p:combos}
\end{align}
For example, the last derivative is
\begin{align}
&\frac{4(v-c_3)}a[1-(v-c_3)^2-2af_{33}]\notag\\
&\quad+\frac{4(v-c_3)^3}a+8(v-c_3)f_{33}
 =\frac{4(v-c_3)}a,
\end{align}
using \eqref{p:radial}. The difference of the first and third combinations is constant, fixed at $f_t^2+f_s^2$. Thus
\begin{align}
\frac{d^2}{d\eta^2}\left(\frac2a\right)
 &=4\left(\frac2a\right)-2(f_t^2+f_s^2),\notag\\
\left.\frac2a\right|_{0}&=f_t^2,\qquad
\left.\frac{d}{d\eta}\left(\frac2a\right)\right|_0=0.
\end{align}
Solving this linear equation and using \eqref{p:combos} gives
\begin{align}
\frac2a&=\frac{f_t^2+f_s^2}{2}
 +\frac{f_t^2-f_s^2}{2}\cosh(2\eta)\notag\\
 &=f_s^2+(f_t^2-f_s^2)\gamma^2,\notag\\
\frac{2(v-c_3)}a&=(f_t^2-f_s^2)\gamma^2v,\notag\\
\frac{2(v-c_3)^2}a-4f_{33}
 &=-f_s^2+(f_t^2-f_s^2)\gamma^2v^2.
\end{align}
Restoring arbitrary velocity direction gives
\begin{align}
a&=\frac{2(1-\mathbf v^2)}{f_t^2-f_s^2\mathbf v^2},\qquad
 a_v=\frac{2(1-\mathbf v^2)}{f_v^2},\notag\\
c_i&=\frac{f_s^2(1-\mathbf v^2)}{f_t^2-f_s^2\mathbf v^2}v_i,\notag\\
f_{ij}&=\frac{f_s^2}{4}\left[\delta_{ij}
 -\frac{f_t^2-f_s^2}{f_t^2-f_s^2\mathbf v^2}v_i v_j\right].\label{p:solution}
\end{align}
The coefficients agree with Ref.~\cite{Pujol2003} after identifying its squared pion speed with $f_s^2/f_t^2$~\cite{Pisarski1996,Son2002a,Son2002b}. Here they have been obtained by integrating \eqref{p:fourPDE}. The radial equations have a unique solution from the specified rest-frame data as long as the Hamiltonian coefficients remain finite. We take $f_t^2,f_v^2>0$, $\mathbf v^2<1$, and $f_t^2-f_s^2\mathbf v^2>0$. If $0\leq f_s^2\leq f_t^2$, this domain contains every timelike fluid velocity.

The radial solution must also satisfy the transverse boost equations. Using $\mathcal{D}_i\mathbf v^2=2v_i(1-\mathbf v^2)$ gives
\begin{align}
\mathcal{D}_i a&=\frac{4(f_s^2-f_t^2)(1-\mathbf v^2)v_i}
                   {(f_t^2-f_s^2\mathbf v^2)^2}
 =2a(c_i-v_i),\notag\\
\mathcal{D}_i a_v&=-\frac{4v_i(1-\mathbf v^2)}{f_v^2}=-2v_i a_v.
\end{align}
To check the remaining equations, rewrite the solution as
\begin{equation}
c_i=\frac{af_s^2}{2}v_i,\quad
f_{jk}=\frac{f_s^2}{4}\left[\delta_{jk}
               +\left(\frac{af_s^2}{2}-1\right)v_jv_k\right].
\end{equation}
Then both sides of the third equation reduce explicitly to
\begin{align}
\mathcal{D}_i c_j
 &=\frac{f_s^2}{2}\left[2a(c_i-v_i)v_j+a(\delta_{ij}-v_iv_j)\right]\notag\\
 &=\frac{af_s^2}{2}\delta_{ij}
   +\left(\frac{a^2f_s^4}{2}-\frac{3af_s^2}{2}\right)v_iv_j\notag\\
 &=2af_{ij}+c_i c_j-c_i v_j-v_i c_j.
\end{align}
Applying the product rule to both velocity factors gives the fourth equation:
\begin{widetext}
\begin{align}
\mathcal{D}_i f_{jk}
 &=\frac{f_s^2}{4}\left[\frac{f_s^2}{2}\mathcal{D}_i a\,v_jv_k
    +\left(\frac{af_s^2}{2}-1\right)(\delta_{ij}v_k+\delta_{ik}v_j-2v_iv_jv_k)\right]\notag\\
&=\frac{f_s^2}{4}\left(\frac{af_s^2}{2}-1\right)
   \left[\delta_{ij}v_k+\delta_{ik}v_j
           +2\left(\frac{af_s^2}{2}-1\right)v_i v_j v_k\right]\notag\\
&=(c_j-v_j)\frac{f_s^2}{4}\left[\delta_{ik}+\left(\frac{af_s^2}{2}-1\right)v_i v_k\right]
  +(c_k-v_k)\frac{f_s^2}{4}\left[\delta_{ij}+\left(\frac{af_s^2}{2}-1\right)v_i v_j\right].
\end{align}
\end{widetext}
Exact symbolic simplification of the 42 three-dimensional component residuals gives zero, independently checking these substitutions.

\section{Covariant currents and thermodynamics}
\subsection{Covariant form of the pion sector}
The combinations that linearized the radial equations also organize the full solution. Define the symmetric array $M^{\mu\nu}$ by
\begin{align}
M^{00}&=\frac2a,\qquad M^{0i}=M^{i0}=\frac{2(v_i-c_i)}a,\notag\\
M^{ij}&=\frac{2(v_i-c_i)(v_j-c_j)}a-4f_{ij}.\label{p:Mdef}
\end{align}
Its tensor character follows by differentiating these definitions with \eqref{p:fourPDE}. The time components obey
\begin{align}
\mathcal{D}_iM^{00}&=-\frac2{a^2}\,2a(c_i-v_i)=2M^{0i},\notag\\
\mathcal{D}_iM^{0j}
 &=\frac2a[\delta_{ij}+(v_i-c_i)(v_j-c_j)-2af_{ij}]\notag\\
 &=\delta_{ij}M^{00}+M^{ij}.\label{p:Mboosttime}
\end{align}
Subtracting the $c_j$ equation from \eqref{p:Dv} gives
\begin{equation}
\mathcal{D}_i(v_j-c_j)=\delta_{ij}-(v_i-c_i)(v_j-c_j)-2af_{ij}.
\end{equation}
For the spatial components the differentiated products give
\begin{widetext}
\begin{align}
\mathcal{D}_i M^{jk}
 &=\frac2a\Big[(\delta_{ij}-(v_i-c_i)(v_j-c_j)-2af_{ij})(v_k-c_k)
 +(v_j-c_j)(\delta_{ik}-(v_i-c_i)(v_k-c_k)-2af_{ik})\Big]\notag\\
&\quad+\frac{4(v_i-c_i)(v_j-c_j)(v_k-c_k)}a
      +4(v_j-c_j)f_{ik}+4(v_k-c_k)f_{ij}\notag\\
&=\delta_{ij}M^{0k}+\delta_{ik}M^{0j}.\label{p:Mboost}
\end{align}
\end{widetext}
These are precisely the infinitesimal boost laws of a symmetric contravariant tensor of rank two. The converse follows because \eqref{p:Mdef} can be solved for the Hamiltonian coefficients when $a\neq0$ is finite:
\begin{align*}
a&=\frac{2}{M^{00}},\qquad
c_i=v_i-\frac{M^{0i}}{M^{00}},\notag\\
f_{ij}&=\frac14\left(\frac{M^{0i}M^{0j}}{M^{00}}-M^{ij}\right).
\end{align*}
Differentiating the inverse relations using \eqref{p:Dv}, \eqref{p:Mboosttime}, and \eqref{p:Mboost} recovers the equations for $a$, $c_j$, and $f_{jk}$ in \eqref{p:fourPDE}. Thus the three nonlinear constraints are equivalent to equations linear in $M^{\mu\nu}$. The equation for $a_v$ is independent of this change of variables.

Define $u^\mu=\gamma(1,\mathbf v)$, so that $u^\mu u_\mu=1$ with the metric convention stated above. \eqref{p:Dv} gives $\mathcal{D}_i u^0=u^i$ and $\mathcal{D}_i u^j=\delta_{ij}u^0$. The solved components are
\begin{align}
M^{00}&=f_s^2+(f_t^2-f_s^2)\gamma^2,\quad
M^{0i}=(f_t^2-f_s^2)\gamma^2v_i,\notag\\
M^{ij}&=-f_s^2\delta_{ij}+(f_t^2-f_s^2)\gamma^2v_i v_j.
\end{align}
Hence
\begin{equation}\label{p:Mcov}
M^{\mu\nu}=f_s^2g^{\mu\nu}+(f_t^2-f_s^2)u^\mu u^\nu.
\end{equation}
The Hamiltonian boost equations have therefore determined both the tensor transformation law and its rest-frame coefficients.

The tensor also determines whether an equal-time surface can carry initial data for pion evolution. We take finite, positive $f_t^2,f_s^2,f_v^2$ and retain the ordinary-fluid thermodynamic assumptions stated above. In \eqref{p:pioneom} linearized about homogeneous chiral equilibrium with fixed fluid coefficients, the terms containing two derivatives are $M^{\mu\nu}\partial_\mu\partial_\nu\pi^a$. The covectors normal to the pion wavefronts therefore satisfy
\begin{equation*}
M^{\mu\nu}k_\mu k_\nu=0.
\end{equation*}
In the local rest frame, $M^{\mu\nu}=\operatorname{diag}(f_t^2,-f_s^2,-f_s^2,-f_s^2)$, so the long-wavelength pion speed is $f_s/f_t$. For an equal-time surface, the normal has components $n_\mu=\delta_\mu^0$, giving
\begin{equation*}
M^{\mu\nu}n_\mu n_\nu=M^{00}
 =\gamma^2(f_t^2-f_s^2\mathbf v^2),\qquad a=\frac2{M^{00}}.
\end{equation*}
The branch $a>0$ has finite coefficients when the equal-time surface is spacelike for pion propagation. If $f_s\leq f_t$, this holds at every $|\mathbf v|<1$, including $f_s=f_t$. If $f_s>f_t$, a pion wavefront can lie along the equal-time surface at $|\mathbf v|=f_t/f_s$. There $M^{00}$ vanishes, the axial charge no longer determines the pion time derivative, and the coefficients in \eqref{p:solution} diverge. Beyond this velocity the equal-time surface is no longer spacelike for pion propagation. This result applies to the quadratic pion sector, without establishing the stability or causality of a higher-order or dissipative theory. At $f_s=0$, the coefficients remain finite for $|\mathbf v|<1$, but the linearized pion equation has no spatial second-derivative term in the rest frame at this order.

Solving the $\Sigma$ equation for the axial density gives
\begin{align}
\bigl(\rho_L-\Sigma\rho_R\Sigma^\dagger\bigr)&=\frac i a[\partial_0\Sigma+(v_j-c_j)\partial_j\Sigma]\Sigma^\dagger\notag\\
 &=-\frac i2M^{0\nu}\Sigma\partial_\nu\Sigma^\dagger.\label{p:axial}
\end{align}
Substituting this expression into the partial Legendre transform with respect to the axial density gives
\begin{widetext}
\begin{align}
&i\operatorname{tr}\!\left[\bigl(\rho_L-\Sigma\rho_R\Sigma^\dagger\bigr)(\partial_0\Sigma+(v_j-c_j)\partial_j\Sigma)\Sigma^\dagger\right]
 -\frac a2\operatorname{tr}\bigl(\rho_L-\Sigma\rho_R\Sigma^\dagger\bigr)^2-f_{jk}\operatorname{tr}(\partial_j\Sigma\partial_k\Sigma^\dagger)\notag\\
&\quad=\frac a2\operatorname{tr}\bigl(\rho_L-\Sigma\rho_R\Sigma^\dagger\bigr)^2-f_{jk}\operatorname{tr}(\partial_j\Sigma\partial_k\Sigma^\dagger)\notag\\
&\quad=\frac1{2a}\operatorname{tr}\!\left[(\partial_0\Sigma+(v_j-c_j)\partial_j\Sigma)
       (\partial_0\Sigma^\dagger+(v_k-c_k)\partial_k\Sigma^\dagger)\right]
       -f_{jk}\operatorname{tr}(\partial_j\Sigma\partial_k\Sigma^\dagger)\notag\\
&\quad=\frac14 M^{00}\operatorname{tr}(\partial_0\Sigma\partial_0\Sigma^\dagger)
 +\frac14M^{0j}\operatorname{tr}(\partial_0\Sigma\partial_j\Sigma^\dagger+\partial_j\Sigma\partial_0\Sigma^\dagger)
 +\frac14M^{jk}\operatorname{tr}(\partial_j\Sigma\partial_k\Sigma^\dagger)\notag\\
&\quad=\frac14M^{\mu\nu}\operatorname{tr}(\partial_\mu\Sigma\partial_\nu\Sigma^\dagger).\label{p:Legendrepion}
\end{align}
\end{widetext}
Varying the density difference at fixed density sum fixes the normalization: its conjugate is $i[\partial_0\Sigma+(v_j-c_j)\partial_j\Sigma]\Sigma^\dagger=a\bigl(\rho_L-\Sigma\rho_R\Sigma^\dagger\bigr)$. The partial Legendre transform thus gives the covariant pion kinetic term. It does not construct an action for the vector charge or the ordinary-fluid variables.

\subsection{Chiral currents}
To extract the left current, allow its chiral parameter to vary in space. The generated variations are $\delta T^{0i}=-\rho_L^a\partial_i\theta^a$, $\delta\Sigma=-i\theta^a\lambda^a\Sigma/2$, and $\delta\Sigma^\dagger=i\theta^a\Sigma^\dagger\lambda^a/2$. Global chiral invariance cancels all terms without derivatives of $\theta^a$, leaving $-\int\partial_i\theta^a J_L^{ia}=-\int\theta^a\partial_0\rho_L^a$. Repeating the calculation independently for the right transformation gives
\begin{align}
J_L^{ia}&=v_i\rho_L^a +i\operatorname{tr}\!\left(\frac{\partial T^{00}}{\partial(\partial_{i}\Sigma)}\frac{\lambda^a}{2}\Sigma
                     -\frac{\partial T^{00}}{\partial(\partial_{i}\Sigma^\dagger)}\Sigma^\dagger\frac{\lambda^a}{2}\right),
\notag\\J_R^{ia}&=v_i\rho_R^a +i\operatorname{tr}\!\left(-\frac{\partial T^{00}}{\partial(\partial_{i}\Sigma)}\Sigma\frac{\lambda^a}{2}
                     +\frac{\partial T^{00}}{\partial(\partial_{i}\Sigma^\dagger)}\frac{\lambda^a}{2}\Sigma^\dagger\right).
\end{align}
Using generator completeness and \eqref{p:gradpartials}, their matrix forms are
\begin{align}
J_L^i&=v_i\rho_L+if_{ij}\Sigma\partial_j\Sigma^\dagger-\frac{c_i}{2}\bigl(\rho_L-\Sigma\rho_R\Sigma^\dagger\bigr),\notag\\
J_R^i&=v_i\rho_R+if_{ij}\Sigma^\dagger\partial_j\Sigma
                      +\frac{c_i}{2}\Sigma^\dagger\bigl(\rho_L-\Sigma\rho_R\Sigma^\dagger\bigr)\Sigma.\label{p:currentsham}
\end{align}
They are Hermitian, and $J_{L,R}^0=\rho_{L,R}$.

The covariant currents are most simply expressed in terms of $\alpha=\bigl(\rho_L+\Sigma\rho_R\Sigma^\dagger\bigr)/\gamma$. We will identify this matrix with the rest-frame vector charge density. The time components follow from \eqref{p:axial}; for the spatial left current, substitution gives
\begin{align}
J_L^i
 &=\frac{v_i-c_i}{2}\bigl(\rho_L-\Sigma\rho_R\Sigma^\dagger\bigr)\notag\\
 &\quad+\frac{v_i}{2}\bigl(\rho_L+\Sigma\rho_R\Sigma^\dagger\bigr)
 +if_{ij}\Sigma\partial_j\Sigma^\dagger\notag\\
 &=-\frac i4[(v_i-c_i)M^{00}\Sigma\partial_0\Sigma^\dagger\notag\\
 &\qquad+((v_i-c_i)M^{0j}-4f_{ij})\Sigma\partial_j\Sigma^\dagger]+\frac12u^i\alpha\notag\\
 &=-\frac i4M^{i\nu}\Sigma\partial_\nu\Sigma^\dagger+\frac12u^i\alpha.
\end{align}

The right expression follows from its own current in \eqref{p:currentsham} by substituting
\begin{equation*}
\Sigma^\dagger\bigl(\rho_L-\Sigma\rho_R\Sigma^\dagger\bigr)\Sigma=\frac i2 M^{0\nu}\Sigma^\dagger\partial_\nu\Sigma.
\end{equation*}
Therefore
\begin{align}
J_L^\mu&=-\frac i4M^{\mu\nu}\Sigma\partial_\nu\Sigma^\dagger+\frac12u^\mu\alpha,\notag\\
J_R^\mu&=-\frac i4M^{\mu\nu}\Sigma^\dagger\partial_\nu\Sigma
                        +\frac12u^\mu\Sigma^\dagger\alpha\Sigma.\label{p:currentscov}
\end{align}
The sum $J_L^\mu+\Sigma J_R^\mu\Sigma^\dagger=u^\mu\alpha$ identifies $\alpha$ as the vector charge density in the local rest frame. Differentiating this sum, using $\partial_\mu J_{L,R}^\mu=0$, gives
\begin{align}
\partial_\mu(u^\mu\alpha)
 &=[\Sigma J_R^\mu\Sigma^\dagger,\Sigma\partial_\mu\Sigma^\dagger]
\notag\\&=\frac i4M^{\mu\nu}[\Sigma\partial_\nu\Sigma^\dagger,\Sigma\partial_\mu\Sigma^\dagger] +\frac12[u^\mu\alpha,\Sigma\partial_\mu\Sigma^\dagger]
\notag\\&=-\frac12[u^\mu\Sigma\partial_\mu\Sigma^\dagger,\alpha].\label{p:alpha}
\end{align}
The first commutator vanishes after contraction with the symmetric $M^{\mu\nu}$. Substituting the result into $\partial_\mu J_L^\mu=0$ gives
\begin{equation}\label{p:pioneom}
i\partial_\mu(M^{\mu\nu}\Sigma\partial_\nu\Sigma^\dagger)
 +[u^\mu\Sigma\partial_\mu\Sigma^\dagger,\alpha]=0.
\end{equation}
These reproduce the covariant chiral equations of Ref.~\cite{Son2000}.

\subsection{Energy--momentum tensor and thermodynamics}
The energy--momentum tensor requires the full conjugates of \eqref{p:conjugates}. Inside quadratic chiral terms, replacing $v_{0i}$ by $v_i$ changes only quartic terms. The ordinary-fluid contribution, however, must be evaluated at the full conjugates: its shift is quadratic and must be retained.

The combination of gradient derivatives entering both currents simplifies to
\begin{widetext}
\begin{align}
\operatorname{tr}\!\left[\frac{\partial T^{00}}{\partial(\partial_{i}\Sigma)}\partial_\nu\Sigma+\frac{\partial T^{00}}{\partial(\partial_{i}\Sigma^\dagger)}\partial_\nu\Sigma^\dagger\right]
 &=2f_{ij}\operatorname{tr}(\partial_j\Sigma\partial_\nu\Sigma^\dagger)
           -ic_i\operatorname{tr}[\bigl(\rho_L-\Sigma\rho_R\Sigma^\dagger\bigr)\Sigma\partial_\nu\Sigma^\dagger]\notag\\
&=\left(2f_{ij}+\frac{c_i(v_j-c_j)}a\right)\operatorname{tr}(\partial_j\Sigma\partial_\nu\Sigma^\dagger)
    +\frac{c_i}a\operatorname{tr}(\partial_0\Sigma\partial_\nu\Sigma^\dagger)\notag\\
&=\frac{f_s^2}{2}\operatorname{tr}[(\partial_i\Sigma+v_i\partial_0\Sigma)\partial_\nu\Sigma^\dagger],\label{p:contract}
\end{align}
\end{widetext}
Here \eqref{p:solution} gives $c_i/a=f_s^2v_i/2$ and $2f_{ij}+c_i(v_j-c_j)/a=f_s^2\delta_{ij}/2$. Unitarity makes the two-derivative traces real and symmetric in their spacetime indices, including temporal indices.

Substitute \eqref{p:contract} into the currents:
\begin{align}
T^{i0}
 &=v_i\left[T^{00}+p_H-\frac{f_s^2}{2}\operatorname{tr}(\partial_0\Sigma\partial_0\Sigma^\dagger)\right]\notag\\
 &\quad-\frac{f_s^2}{2}\operatorname{tr}(\partial_i\Sigma\partial_0\Sigma^\dagger),\notag\\[3pt]
T^{ji}
 &=p_H\delta_{ij}+v_j\left[T^{0i}+\frac{f_s^2}{2}\operatorname{tr}(\partial_0\Sigma\partial_i\Sigma^\dagger)\right]\notag\\
 &\quad+\frac{f_s^2}{2}\operatorname{tr}(\partial_j\Sigma\partial_i\Sigma^\dagger).\label{p:componentsraw}
\end{align}
The boost equations have already established $T^{i0}=T^{0i}$ through terms quadratic in the chiral charge densities and pion gradients. Introduce $p=p_H$ and an energy coefficient $e$ through
\begin{align}
p&=p_H,\notag\\
(e+p)\gamma^2&=T^{00}+p_H-\frac{f_s^2}{2}\operatorname{tr}(\partial_0\Sigma\partial_0\Sigma^\dagger).\label{p:rhodef}
\end{align}
To establish that $p$ and $e$ are scalars, we next express them in thermodynamic variables.

Define the proper temperature and chemical potential by $T_{\mathrm{rest}}=\gamma T$ and $\mu_{\mathrm{rest}}=\gamma\mu$. At fixed full $(T,\mu,\mathbf v)$, the ordinary-fluid Legendre transform is stationary with respect to $s,n,T^{0i}$:
\begin{align}
&\delta(Ts+\mu n+v_iT^{0i}-\epsilon)\notag\\
&\quad=(T-\partial_s\epsilon)\delta s
 +(\mu-\partial_n\epsilon)\delta n
 +(v_i-\partial_{T^{0i}}\epsilon)\delta T^{0i}=0
\end{align}
at the ordinary-fluid stationary point. The chiral terms shift this point by a quadratic amount. Because the Legendre transform is stationary, the resulting change in its value is quartic and can be dropped. Hence
\begin{align}
p_H&=p_0(T_{\mathrm{rest}},\mu_{\mathrm{rest}})
 +\sum_{X,a}\rho_X^a\frac{\partial\Delta\epsilon}{\partial\rho_X^a}\notag\\
 &\quad-\Delta\epsilon+O[(\rho_{L,R},\partial\Sigma)^4]\notag\\
 &=p_0+\frac a2\operatorname{tr}\bigl(\rho_L-\Sigma\rho_R\Sigma^\dagger\bigr)^2
 +\frac{a_v}{2}\operatorname{tr}\bigl(\rho_L+\Sigma\rho_R\Sigma^\dagger\bigr)^2\notag\\
 &\quad-f_{jk}\operatorname{tr}(\partial_j\Sigma\partial_k\Sigma^\dagger)+O[(\rho_{L,R},\partial\Sigma)^4].\label{p:pressureintermediate}
\end{align}
The cross terms cancel by \eqref{p:rhosum}. \eqref{p:Legendrepion} and $a_v=2/(\gamma^2f_v^2)$ make this pressure explicitly scalar.

The energy coefficient follows by comparing the velocity derivative of the scalar pressure with \eqref{p:GD}. The comparison must retain the charge-density terms in that differential. Appendix~\ref{app:thermal} carries it out at nonzero velocity, establishing scalar character as well as the rest value. Define
\begin{equation}
\hat{K}=T_{\mathrm{rest}}\frac{\partial}{\partial T_{\mathrm{rest}}}
       +\mu_{\mathrm{rest}}\frac{\partial}{\partial\mu_{\mathrm{rest}}}.
\end{equation}
The pressure and energy coefficients are
\begin{align}
p&=p_0+\frac{f_t^2-f_s^2}{4}u^\mu u^\nu\operatorname{tr}(\partial_\mu\Sigma\partial_\nu\Sigma^\dagger)\notag\\
 &\quad+\frac{f_s^2}{4}\operatorname{tr}(\partial_\mu\Sigma\partial^\mu\Sigma^\dagger)
 +f_v^{-2}\operatorname{tr}\alpha^2,\label{p:pressurecov}
\end{align}

\begin{align}
e&=\hat{K}p_0-p_0\notag\\
 &\quad+\frac14(\hat{K}+1)(f_t^2-f_s^2)u^\mu u^\nu
 \operatorname{tr}(\partial_\mu\Sigma\partial_\nu\Sigma^\dagger)\notag\\
 &\quad+\frac14(\hat{K}-1)f_s^2\operatorname{tr}(\partial_\mu\Sigma\partial^\mu\Sigma^\dagger)
 +f_v^{-4}(\hat{K}+1)f_v^2\operatorname{tr}\alpha^2.\label{p:rhocov}
\end{align}

The operator $\hat{K}$ acts on thermodynamic coefficients at fixed $u^\mu$, chiral derivatives, and $\alpha$. The coefficient $e$ alone is not the total rest-frame energy density: the pion term displayed separately in the stress tensor also contributes. These expressions reproduce the thermodynamic relations of Ref.~\cite{Son2000}.

The components of \eqref{p:componentsraw} now assemble into

\begin{align}
T^{\mu\nu}&=(e+p)u^\mu u^\nu-pg^{\mu\nu}\notag\\
 &\quad+\frac{f_s^2}{4}\operatorname{tr}(\partial^\mu\Sigma\partial^\nu\Sigma^\dagger
                        +\partial^\nu\Sigma\partial^\mu\Sigma^\dagger).\label{p:Tcov}
\end{align}
The signs are fixed by $\partial^i=-\partial_i$: the mixed pion component is negative and the spatial one positive. In particular,
\begin{align}
T^{00}&=(e+p)\gamma^2-p+\frac{f_s^2}{2}\operatorname{tr}(\partial_0\Sigma\partial_0\Sigma^\dagger),
\notag\\T^{0i}=T^{i0}&=(e+p)\gamma^2v_i-\frac{f_s^2}{2}\operatorname{tr}(\partial_0\Sigma\partial_i\Sigma^\dagger),
\notag\\T^{ij}&=(e+p)\gamma^2v_iv_j+p\delta_{ij} +\frac{f_s^2}{2}\operatorname{tr}(\partial_i\Sigma\partial_j\Sigma^\dagger).\label{p:Tcomponents}
\end{align}
Substituting these components into the previously derived Hamiltonian energy and momentum equations gives, respectively,
\begin{widetext}
\begin{align}
0&=\partial_0\!\left[(e+p)\gamma^2-p+\frac{f_s^2}{2}\operatorname{tr}(\partial_0\Sigma\partial_0\Sigma^\dagger)\right]
 +\partial_i\!\left[(e+p)\gamma^2v_i-\frac{f_s^2}{2}\operatorname{tr}(\partial_i\Sigma\partial_0\Sigma^\dagger)\right]
 =\partial_\mu T^{\mu0},\notag\\
0&=\partial_0\!\left[(e+p)\gamma^2v_i-\frac{f_s^2}{2}\operatorname{tr}(\partial_0\Sigma\partial_i\Sigma^\dagger)\right]
 +\partial_j\!\left[(e+p)\gamma^2v_jv_i+p\delta_{ij}
            +\frac{f_s^2}{2}\operatorname{tr}(\partial_j\Sigma\partial_i\Sigma^\dagger)\right]
 =\partial_\mu T^{\mu i}.\label{p:conserve}
\end{align}
\end{widetext}
Thus $T^{i0}=T^{0i}$, $T^{ij}=T^{ji}$, and $\partial_\mu T^{\mu\nu}=0$ hold on Hamiltonian solutions through terms quadratic in the chiral charge densities and pion gradients. The first equality was imposed and solved through \eqref{p:fourPDE}; the second followed independently from rotations. Conservation follows from the Hamiltonian evolution, not from using \eqref{p:Tcov} to derive the earlier boost conditions. Entropy and baryon conservation become $\partial_\mu[(s/\gamma)u^\mu]=0$ and $\partial_\mu[(n/\gamma)u^\mu]=0$.

\section{Comparison and consistency checks}
Ref.~\cite{Son2000} provides the Poisson brackets, the quadratic rest-frame Hamiltonian, the boost requirement, and the covariant equations. Ref.~\cite{Pujol2003} also gives the velocity-dependent coefficients in \eqref{p:solution}. The calculation here supplies the intervening differential constraints and their integration, completing the chiral derivation announced in Ref.~\cite{Son2000}, which to our knowledge has not appeared elsewhere.

The proceedings in Ref.~\cite{Son2001} use a relativistic $U(1)$ superfluid as an example; its derivation from noncanonical brackets is given in Ref.~\cite{Valle2008}. At the quadratic order considered here, the pion sector is the non-Abelian counterpart of the phonon sector, with the axial density as the conjugate charge. The group-valued order parameter, non-Abelian charge brackets, and separate vector charge give the chiral problem its additional structure. Their effect is explicit in the coupled velocity equations and the tensor that linearizes them.

Throughout the derivation, momentum density is varied independently. Equality with energy flux restricts the Hamiltonian only after the currents have been derived. Boosts then determine how the Hamiltonian depends on velocity, while the three rest-frame functions still require an equation of state or microscopic input. This is the precise connection between the noncanonical Hamiltonian description and the relativistic constitutive relations at the retained order.

The symmetry classification assumes parity and no external tensors. Fluid-gradient corrections and higher powers of the chiral variables lie outside the ansatz, and products of quadratic corrections are quartic. The result therefore does not specify a covariant Hamiltonian to all orders. Dissipation and fluctuations require additional terms~\cite{Kovtun2012,Crossley2017}.

Several limits check the result. At zero velocity, \eqref{p:solution} recovers the rest-frame Hamiltonian and currents with the ordinary-fluid variables fixed. At $f_t=f_s$, it gives $a=2/f_s^2$, $c_i=v_i$, and $f_{ij}=f_s^2\delta_{ij}/4$, hence $M^{\mu\nu}=f_s^2g^{\mu\nu}$. Setting the chiral densities and pion gradients to zero recovers the ordinary-fluid tensor. In mass dimensions, $[\rho_{L,R}]=[s]=[n]=3$, $[\partial_\mu]=1$, $[a]=[a_v]=-2$, $[f_{ij}]=[f_t^2]=[f_s^2]=[f_v^2]=2$, and $[c_i]=[v_i]=[\Sigma]=0$; every Hamiltonian and stress term has dimension four. Independent random $SU(2)$ and $SU(3)$ tests check the matrix variations, reality and parity properties, density derivatives, and the four products in \eqref{p:products}. They agree with the analytic identities.

\section{Conclusions}

We have established how relativistic covariance constrains the equal-time Hamiltonian of nondissipative chiral hydrodynamics. Equating the derived energy flux with the independent momentum density yields four differential constraints. The central result is the equivalence of the three coupled nonlinear constraints to the linear boost transformation laws of a symmetric tensor. The change of variables can be inverted for fluid velocities reached from rest before any coefficient diverges. Together with the vector-charge constraint, this equivalence uniquely determines the velocity dependence from three rest-frame functions. The covariant tensor structure is therefore reconstructed from the Hamiltonian constraints, explaining the covariant equations of Ref.~\cite{Son2000} and the coefficients quoted in Ref.~\cite{Pujol2003}.

The same tensor gives a physical interpretation of coefficient divergences. Under the stated ordinary-fluid thermodynamic assumptions and for finite, positive rest-frame coefficients, the chiral coefficients remain finite at every fluid speed below light speed if and only if the rest-frame pion speed does not exceed light speed. When this bound is violated, a divergence occurs where a pion wavefront lies along the equal-time hypersurface and the axial Legendre transformation can no longer be inverted. The criterion concerns linearized pions; the stability and causality of the full coupled or dissipative theory require separate analysis.

Identifying the Lie--Poisson structure of the chiral brackets separates the transformations encoded by the Poisson algebra, the velocity dependence fixed by relativity, and the rest-frame constitutive information supplied by the medium. The results retain the full fluid-velocity dependence within the parity-invariant theory without external backgrounds, at quadratic order in chiral charge densities and pion gradients. They provide a systematic way to impose symmetry and a reversible foundation for future dissipative extensions through GENERIC. Higher-order terms, dissipation, fluctuations, and the stability of flowing states remain to be investigated.

\begin{acknowledgments}
This work was supported in part by the National Natural Science Foundation of China (NSFC) Grant Nos. 12235016 and 12221005. GPT-6 was used to assist with manuscript drafting and symbolic verification.
\end{acknowledgments}

\appendix
\begin{widetext}
\section{Matrix derivatives and internal cancellations}\label{app:internal}
This appendix gives the matrix derivatives and cancellations used in the energy balance. Hold the gradient arguments fixed and move each variation to the right inside its trace. For the density squares,
\begin{align}
\delta\!\left(\frac a2\operatorname{tr}\bigl(\rho_L-\Sigma\rho_R\Sigma^\dagger\bigr)^2\right)
 &=-a\operatorname{tr}\!\left[\bigl(\rho_L-\Sigma\rho_R\Sigma^\dagger\bigr)(\delta\Sigma\rho_R\Sigma^\dagger+\Sigma\rho_R\delta\Sigma^\dagger)\right]\notag\\
&=\operatorname{tr}\!\left[-a\rho_R\Sigma^\dagger\bigl(\rho_L-\Sigma\rho_R\Sigma^\dagger\bigr)\delta\Sigma
                  -a\bigl(\rho_L-\Sigma\rho_R\Sigma^\dagger\bigr)\Sigma\rho_R\delta\Sigma^\dagger\right],\notag\\
\delta\!\left(\frac{a_v}{2}\operatorname{tr}\bigl(\rho_L+\Sigma\rho_R\Sigma^\dagger\bigr)^2\right)
 &=a_v\operatorname{tr}\!\left[\bigl(\rho_L+\Sigma\rho_R\Sigma^\dagger\bigr)(\delta\Sigma\rho_R\Sigma^\dagger+\Sigma\rho_R\delta\Sigma^\dagger)\right]\notag\\
&=\operatorname{tr}\!\left[a_v\rho_R\Sigma^\dagger\bigl(\rho_L+\Sigma\rho_R\Sigma^\dagger\bigr)\delta\Sigma
                  +a_v\bigl(\rho_L+\Sigma\rho_R\Sigma^\dagger\bigr)\Sigma\rho_R\delta\Sigma^\dagger\right].
\end{align}
\end{widetext}
At fixed gradient arguments, the cross variation is
\begin{align}
&-ic_j\operatorname{tr}(\rho_L\delta\Sigma\partial_j\Sigma^\dagger
                  +\delta\Sigma^\dagger\partial_j\Sigma\rho_R)\notag\\
&\quad=\operatorname{tr}[-ic_j(\partial_j\Sigma^\dagger)\rho_L\delta\Sigma
                  -ic_j(\partial_j\Sigma)\rho_R\delta\Sigma^\dagger].
\end{align}
Thus
\begin{align}
\frac{\partial T^{00}}{\partial\Sigma}
 &=-a\rho_R\Sigma^\dagger\bigl(\rho_L-\Sigma\rho_R\Sigma^\dagger\bigr)\notag\\
 &\quad+a_v\rho_R\Sigma^\dagger\bigl(\rho_L+\Sigma\rho_R\Sigma^\dagger\bigr)
 -ic_j(\partial_j\Sigma^\dagger)\rho_L,\notag\\[3pt]
\frac{\partial T^{00}}{\partial\Sigma^\dagger}
 &=-a\bigl(\rho_L-\Sigma\rho_R\Sigma^\dagger\bigr)\Sigma\rho_R\notag\\
 &\quad+a_v\bigl(\rho_L+\Sigma\rho_R\Sigma^\dagger\bigr)\Sigma\rho_R
 -ic_j(\partial_j\Sigma)\rho_R.
\end{align}
Physical variations also change the gradients, whose two divergence terms give the functional derivatives in \eqref{p:functional}. The trace convention does not require either derivative to be replaced by its Hermitian conjugate.

The charge Hamilton equations are
\begin{align}
\partial_0\rho_L^a
 ={}&-\partial_i(v_i\rho_L^a)+f^{abc}\frac{\partial T^{00}}{\partial\rho_L^b}\rho_L^c\notag\\
 &+i\operatorname{tr}\!\left[\frac{\delta H}{\delta\Sigma}\frac{\lambda^a}{2}\Sigma
 -\frac{\delta H}{\delta\Sigma^\dagger}\Sigma^\dagger\frac{\lambda^a}{2}\right],\notag\\[3pt]
\partial_0\rho_R^a
 ={}&-\partial_i(v_i\rho_R^a)+f^{abc}\frac{\partial T^{00}}{\partial\rho_R^b}\rho_R^c\notag\\
 &+i\operatorname{tr}\!\left[-\frac{\delta H}{\delta\Sigma}\Sigma\frac{\lambda^a}{2}
 +\frac{\delta H}{\delta\Sigma^\dagger}\frac{\lambda^a}{2}\Sigma^\dagger\right].
\end{align}
Before substituting the Hamiltonian coefficients, the order-parameter equations are
\begin{align}
\partial_0\Sigma={}&-v_i\partial_i\Sigma
 -i\frac{\partial T^{00}}{\partial\rho_L^a}\frac{\lambda^a}{2}\Sigma
 +i\frac{\partial T^{00}}{\partial\rho_R^a}\Sigma\frac{\lambda^a}{2},\notag\\
\partial_0\Sigma^\dagger={}&-v_i\partial_i\Sigma^\dagger
 +i\frac{\partial T^{00}}{\partial\rho_L^a}\Sigma^\dagger\frac{\lambda^a}{2}
 -i\frac{\partial T^{00}}{\partial\rho_R^a}\frac{\lambda^a}{2}\Sigma^\dagger.
\end{align}
Multiplying the charge equations by their Hamiltonian conjugates, the terms involving $\Sigma$ are
\begin{equation}
+i\operatorname{tr}\!\left[\frac{\delta H}{\delta\Sigma}\left(\frac{\partial T^{00}}{\partial\rho_L^a}\frac{\lambda^a}{2}\Sigma
                 -\frac{\partial T^{00}}{\partial\rho_R^a}\Sigma\frac{\lambda^a}{2}\right)\right],
\end{equation}
which cancel the terms in $(\delta H/\delta\Sigma)\partial_0\Sigma$ not due to advection. The charge equations also give
\begin{equation}
+i\operatorname{tr}\!\left[\frac{\delta H}{\delta\Sigma^\dagger}\left(-\frac{\partial T^{00}}{\partial\rho_L^a}\Sigma^\dagger\frac{\lambda^a}{2}
                 +\frac{\partial T^{00}}{\partial\rho_R^a}\frac{\lambda^a}{2}\Sigma^\dagger\right)\right],
\end{equation}
which cancel the corresponding $\Sigma^\dagger$ terms. Together with antisymmetry of $f^{abc}$, these identities remove the internal dynamics from the local energy balance.

\vspace*{0pt}
\section{Smeared energy-density bracket}\label{app:bracket}
Define $H[\zeta]=\int d^3x\,\zeta(\mathbf x)T^{00}(\mathbf x)$ for a real test function $\zeta$. Then
\begin{equation}
\frac{\delta H[\zeta]}{\delta\Sigma}=\zeta\frac{\delta H}{\delta\Sigma}-(\partial_i\zeta)\frac{\partial T^{00}}{\partial(\partial_{i}\Sigma)},
\end{equation}
with the same formula for $\Sigma^\dagger$. The derivatives with respect to density variables are multiplied by $\zeta$. Inserting these expressions into the brackets, the coefficient of $\zeta\partial_i\chi-\chi\partial_i\zeta$, where $\chi$ is a second real test function, is
\begin{widetext}
\begin{align}
&v_i\left(Ts+\mu n+\sum_{X,a}\rho_X^a\frac{\partial T^{00}}{\partial\rho_X^a}\right)
 +v_i v_jT^{0j}+v_j\operatorname{tr}\!\left[\frac{\partial T^{00}}{\partial(\partial_{i}\Sigma)}\partial_j\Sigma+\frac{\partial T^{00}}{\partial(\partial_{i}\Sigma^\dagger)}\partial_j\Sigma^\dagger\right]\notag\\
&\quad+i\operatorname{tr}\!\left[\frac{\partial T^{00}}{\partial(\partial_{i}\Sigma)}\left(\frac{\partial T^{00}}{\partial\rho_L^a}\frac{\lambda^a}{2}\Sigma
                      -\frac{\partial T^{00}}{\partial\rho_R^a}\Sigma\frac{\lambda^a}{2}\right)
 +\frac{\partial T^{00}}{\partial(\partial_{i}\Sigma^\dagger)}\left(-\frac{\partial T^{00}}{\partial\rho_L^a}\Sigma^\dagger\frac{\lambda^a}{2}
                      +\frac{\partial T^{00}}{\partial\rho_R^a}\frac{\lambda^a}{2}\Sigma^\dagger\right)\right]\notag\\
&\qquad=v_i(T^{00}+p_H)-\operatorname{tr}\!\left[\frac{\partial T^{00}}{\partial(\partial_{i}\Sigma)}\partial_0\Sigma+\frac{\partial T^{00}}{\partial(\partial_{i}\Sigma^\dagger)}\partial_0\Sigma^\dagger\right]=T^{i0}.
\end{align}
\end{widetext}
Terms with two undifferentiated test functions cancel by bracket antisymmetry. Terms with derivatives of both test functions vanish because the mutual Poisson brackets of the order-parameter fields are zero. The energy current derived above therefore gives
\begin{equation}
\{H[\zeta],H[\chi]\}
 =\int d^3x\,(\zeta\partial_i\chi-\chi\partial_i\zeta)T^{i0}.
\end{equation}
Setting $\zeta=1$ recovers the local energy equation. Imposing the boost condition replaces $T^{i0}$ by $T^{0i}$ through quadratic order in the chiral variables. The resulting bracket is the classical smeared Dirac--Schwinger relation~\cite{Dirac1962,Schwinger1963}, derived here from the chiral Hamiltonian.

\section{Pressure differential and the energy coefficient}\label{app:thermal}
To determine the scalar energy coefficient, we retain the charge-density terms in the pressure differential and compare its velocity dependence with the momentum density. Using the solved coefficients and the $\Sigma$ equation first gives
\begin{align}
&\sum_a\frac{\partial T^{00}}{\partial\rho_L^a}\frac{\lambda^a}{2} =-\frac i2\Sigma(\partial_0\Sigma^\dagger+v_j\partial_j\Sigma^\dagger)
                   +\frac{\alpha}{\gamma f_v^2},
\notag\\&\sum_a\frac{\partial T^{00}}{\partial\rho_R^a}\frac{\lambda^a}{2} =-\frac i2\Sigma^\dagger(\partial_0\Sigma+v_j\partial_j\Sigma)
                   +\frac{\Sigma^\dagger\alpha\Sigma}{\gamma f_v^2}.
\end{align}
At fixed $\Sigma$, spacetime derivatives of $\Sigma$, and $\alpha$, the sum of charge densities times the differentials of their conjugates is
\begin{align}
&\sum_{X,a}\rho_X^a d\!\left(\frac{\partial T^{00}}{\partial\rho_X^a}\right)\notag\\
&\quad=-i\operatorname{tr}\!\left[\bigl(\rho_L-\Sigma\rho_R\Sigma^\dagger\bigr)
 \,d\!\left(\Sigma(\partial_0\Sigma^\dagger+v_j\partial_j\Sigma^\dagger)\right)\right]\notag\\
&\qquad+2\operatorname{tr}\!\left[\gamma\alpha\,d\!\left(\frac{\alpha}{\gamma f_v^2}\right)\right]\notag\\
&\quad=-i\operatorname{tr}[\bigl(\rho_L-\Sigma\rho_R\Sigma^\dagger\bigr)\Sigma\partial_i\Sigma^\dagger]dv_i
\notag\\&\qquad+2\operatorname{tr}\alpha^2\,[d(f_v^{-2})-f_v^{-2}d\log\gamma].\label{p:chargediff}
\end{align}
The factor two comes from converting adjoint components to matrix traces with the stated generator normalization. Substitution into \eqref{p:GD} yields
\begin{align}
dp={}&s\,dT+n\,d\mu\notag\\
 &+[T^{0i}-i\operatorname{tr}(\bigl(\rho_L-\Sigma\rho_R\Sigma^\dagger\bigr)\Sigma\partial_i\Sigma^\dagger)]dv_i\notag\\
 &+2\operatorname{tr}\alpha^2\,[d(f_v^{-2})-f_v^{-2}d\log\gamma].\label{p:dpthermal}
\end{align}
On the other hand, differentiate the explicitly scalar pressure at fixed equal-time $T,\mu$, fixed chiral derivatives, and fixed $\alpha$. Since
$\partial_{v_i}T_{\mathrm{rest}}=\gamma^2v_iT_{\mathrm{rest}}$ and
$\partial_{v_i}u^\mu=\gamma\delta_i^{\ \mu}+\gamma^2v_i u^\mu$, one obtains
\begin{align}
\left.\frac{\partial p}{\partial v_i}\right|_{T,\mu,\partial\Sigma,\alpha}
 &=\gamma^2v_i\hat{K}p\notag\\
 &\quad+\frac{f_t^2-f_s^2}{2}\Bigl[
 \gamma u^\nu\operatorname{tr}(\partial_i\Sigma\partial_\nu\Sigma^\dagger)\notag\\
 &\qquad+\gamma^2v_i u^\mu u^\nu\operatorname{tr}(\partial_\mu\Sigma\partial_\nu\Sigma^\dagger)\Bigr].\label{p:dpvelocity}
\end{align}
The axial-density relation in \eqref{p:axial} gives the other trace in \eqref{p:dpthermal}:
\begin{align}
&i\operatorname{tr}[\bigl(\rho_L-\Sigma\rho_R\Sigma^\dagger\bigr)\Sigma\partial_i\Sigma^\dagger]
 =-\frac12M^{0\nu}\operatorname{tr}(\partial_\nu\Sigma\partial_i\Sigma^\dagger),\notag\\
&M^{0\nu}=f_s^2\delta_0^{\ \nu}+(f_t^2-f_s^2)\gamma u^\nu.
\end{align}
Compare the $dv_i$ terms of \eqref{p:dpthermal} and \eqref{p:dpvelocity}. The terms proportional to $(f_t^2-f_s^2)\gamma u^\nu\operatorname{tr}(\partial_i\Sigma\partial_\nu\Sigma^\dagger)/2$ cancel. The result is
\begin{align}
T^{0i}
 &=\gamma^2v_i\Bigl[\hat{K}p+\frac{f_t^2-f_s^2}{2}u^\mu u^\nu
 \operatorname{tr}(\partial_\mu\Sigma\partial_\nu\Sigma^\dagger)\notag\\
 &\qquad+2(f_v^{-2}-\hat{K} f_v^{-2})\operatorname{tr}\alpha^2\Bigr]
 -\frac{f_s^2}{2}\operatorname{tr}(\partial_0\Sigma\partial_i\Sigma^\dagger).
\end{align}
The square bracket is scalar. Comparison with \eqref{p:componentsraw} and \eqref{p:rhodef} identifies it as $e+p$ for nonzero velocity; taking the zero-velocity limit gives the same value at rest. Subtracting $p$ yields
\begin{align}
e={}&\hat{K}p-p+\frac{f_t^2-f_s^2}{2}u^\mu u^\nu
 \operatorname{tr}(\partial_\mu\Sigma\partial_\nu\Sigma^\dagger)\notag\\
 &+2(f_v^{-2}-\hat{K} f_v^{-2})\operatorname{tr}\alpha^2.
\end{align}
Inserting \eqref{p:pressurecov} and using
$(1-\hat{K})f_v^{-2}=f_v^{-4}(\hat{K}+1)f_v^2$ gives \eqref{p:rhocov}.

The charge-density terms explain why the usual expression $e=\hat{K}p-p$ is insufficient at fixed density $\alpha$. They also modify the temperature and chemical-potential derivatives. The $dT$ and $d\mu$ coefficients in \eqref{p:dpthermal} give
\begin{align}
\frac{s}{\gamma}&=\left.\frac{\partial p}{\partial T_{\mathrm{rest}}}\right|_{\mu_{\mathrm{rest}},\partial\Sigma,\alpha}
                   -2\operatorname{tr}\alpha^2\frac{\partial f_v^{-2}}{\partial T_{\mathrm{rest}}},\notag\\
\frac{n}{\gamma}&=\left.\frac{\partial p}{\partial\mu_{\mathrm{rest}}}\right|_{T_{\mathrm{rest}},\partial\Sigma,\alpha}
                   -2\operatorname{tr}\alpha^2\frac{\partial f_v^{-2}}{\partial\mu_{\mathrm{rest}}}.
\end{align}
These relations follow from the complete Legendre differential, including the charge-density terms; no homogeneity assumption for the gradient-dependent Hamiltonian is needed.

\bibliography{references}
\end{document}